\documentclass[a4paper,11pt]{article}
\pdfoutput=1 

\usepackage{jheppub} 
\usepackage[T1]{fontenc} 
\usepackage{amsmath}

\makeatletter
\def\@fpheader{\relax}
\makeatother
\renewcommand\footnoterule{\kern-2pt \hrule width 6in \kern 2.6pt}
\title{\boldmath Schwinger type pair production in Non-SuSy AdS/CFT}

\author{Udit Narayan Chowdhury}

\affiliation{Saha Institute of Nuclear Physics, Block-AF, Sector-1, Salt Lake.\\
 Kolkata 700064, India
 
 and

Homi Bhabha National Institute, Training School Complex, 
Anushakti Nagar.\\
Mumbai 400094, India}

\emailAdd{udit.chowdhury@saha.ac.in, chowdhury.udit@gmail.com}

\abstract{We study pair production of particles in presence of an external electric field in a large N Non-supersymmetric Yang Mills theory using the holographic duality. The dual geometry we consider
is asymptotically AdS
and is effectively parametrized by two parameters, $u_0$ and $-\sqrt{\frac{5}{2}}<\delta \leqslant 0$ both of which can be related to the effective mass of quark/antiquark for Non-supersymmetric theories.
We numerically calculate
the inter-quark potential profile and the effective potential to study pair production and analytically find out the threshold electric field beyond which one gets catastrophic pair creation by studying 
rectangular Wilson loops using the holographic method. We also find out the critical electric field from DBI analysis of a probe brane. Our initial investigations
reveal, that the critical electric field necessary for spontaneous pair production increases or decreases w.r.t. its Non-Supersymmetric value depending on the parameter $\delta$. Ultimately we find out 
the pair production rate of particles in presence of an external electric field by evaluating circular Wilson loops using 
perturbative methods. From the later investigation we note the resemblance with our earlier prediction. However, we also see that for and below another certain value of the parameter $\delta$ the pair
production rate
of particle/antiparticle pairs blows up as the external electric field is taken to zero. We thus
infer that the vacuum of the Non-SuSy gauge theory is unstable for a range of Non-supersymmetric parameter $\delta$ and that the geometry/Non-SUSY field theory under consideration has quite different
characteristics 
than earlier reported.}

\begin{document} 
\maketitle
\flushbottom

\section{Introduction}
For the last few decades the AdS/CFT correspondence \cite{Maldacena:1997re}\cite{Witten:1998qj}\cite{Aharony:1999ti}\cite{Ammon:2015wua} (relating $\mathcal{N}=4$ Super-Conformal Yang-Mills in
4 spacetime dimensions to quantum gravity in asymptotic $AdS_5 \otimes S^5$ spaces) and some of its modifications
is one  of the exemplary ideas in theoretical physics. AdS/CFT chiefly comes from black hole thermodynamics\cite{Townsend:1997ku} and type IIB string theory \cite{Blumenhagen:2013fgp} is
thus inherently supersymmetric in nature. This is a strong-weak duality
meaning, strong coupling in the field theory side corresponds to weak coupling in the Quantum Gravity side and vice-versa. However even after so many years, no trace of 
supersymmetry has been found by experiments and again conformal symmetry is not found quite much in nature. Thus it is necessary to formulate a modification of AdS/CFT without supersymmetry and
conformal symmetry yet respecting
its string theory/supergravity origins. Such a solution is obtained in \cite{Nayek:2016hsi}\cite{Constable:1999ch}. This solution for D3 branes has two parameters $\delta$ and $u_0$ (i.e. the field 
theory dual
is a two parameter deformation over usual $\mathcal{N}=4$ Super Yang Mills)and
has certain features which makes its an attractive dual for large N QCD studies via 
holography \cite{Nayek:2018aei}. Too the best of our knowledge, an explicit interpretation of these two parameters in terms of dual field theory measurables lacks till date.   It had been
previously reported that \cite{Nayek:2016hsi}\cite{Constable:1999ch}\cite{Nayek:2018aei}  those include, running coupling + confinement in the infrared
and absence of both SUSY and conformal symmetry (thus the 
two parameter deformation commented on above violates both SUSY and conformal symmetry). Part of what makes AdS/CFT alluring is that when the field theory coupling is high the corresponding
coupling in 
the quantum gravity side is low and thus we are left with classical gravity which is easily computable.
\\
\\
The coupling constant in field theory is usually used as a perturbative parameter and observables are expressed in a series w.r.t. this parameter, this is called perturbative field theory. However
there are quite some effects in
quantum field theory which cannot be explained as such i.e. non-perturbative effects. Amongst them the Schwinger Effect stands its ground. The vacuum of QED or any gauge theory interacting with
charged matter is
full of virtual particles and antiparticles (henceforth $q\bar{q}$). In presence of an external electric field/gauge field this particles get the required energy and become real particles. There is
no magic involved in this. In realistic situations the energy of the the real $q\bar{q}$ pairs is obtained from the electric field. Schwinger calculated \cite{PhysRev.82.664} the pair production rate 
for this process in U(1)
gauge theory and
obtained,
\begin{align}
 \Gamma=\frac{(eE)^{3}}{(2\pi)^{3}}e^{\frac{-\pi m^{2}}{eE}}
\end{align}
The exponential suppression hints that pair productions can be modeled as a tunneling process. Assuming that the virtual $q\bar{q}$ pair has a separation x the potential on a virtual quark in presence on an 
external electric field is given as 
\begin{align}
 \mathcal{V}_{eff}=-\frac{\alpha}{x}-eEx+2m
\end{align}
Imagine this to be the potential barrier though which the $q\bar{q}$ pairs tunnel out in the opposite direction and become real. For  E$<\frac{m^2}{e\alpha}$ there exists two zero points
of the
potential and $\mathcal{V}_{eff}$ is positive for intermediate values of x. That means there is a potential barrier and quarks have to tunnel out through them justifying the exponential factor stated above.
However for
E$>\frac{m^2}{e\alpha}$ the potential becomes negative all along and stops putting up a
potential barrier, indicating a catastrophic instability of vacuum where the $q\bar{q}$ are produced spontaneously. The value of electric field for which the potential stops putting up a tunneling barrier is called 
"critical/threshold electric field" $E_c$.
\\
\\The Schwinger effect in holographic setting was first calculated in \cite{Semenoff:2011ng} (see \cite{Gorsky:2001up} for an even earlier work) wherein the pair modified pair production rate was found to be 
\begin{align}
 \Gamma \sim \text{exp}\Big[-\frac{\sqrt{\lambda}}{2}\Big(\sqrt{\frac{E_c}{E}}-\sqrt{\frac{E}{E_c}}\Big)^2\Big]\;\;\;\;\;;\;\;\;\;\;E_c=\frac{2\pi m^2}{\sqrt{\lambda}}
\end{align}
This formula matches with the one above for low electric field (much lower than $E_c$). For field much higher than $E_c$ we don't see a exponential suppression anymore hinting at catastrophic decay. The chief idea of
this work was to place the probe brane at a finite position unlike what is done usually (placing the probe brane at the conformal boundary of AdS) and then to calculate the circular Wilson 
loop. Another approach was pioneered in \cite{Sato:2013iua} which calculated the rectangular Wilson loop for virtual $q\bar{q}$ pair and relate it to inter-quark potential and then 
find the critical
electric field from the same. Holographic Schwinger effect for confining gauge theories have also been studied in literature\cite{Sato:2013dwa}\cite{Kawai:2015mha}, and the confinement manifest itself in the presence of another \textit{"threshold" electric field,
below which pair production doesn't
happen at all}. In this work we want to study the Schwinger effect for Non-Supersymmetric gauge theories via holographic methods using both of this methods, our chief interest being twofold. On
one hand we like to see
 the theoretical effect absence of supersymmetry yields on the value of critical electric field at least for large N Yang Mills theories(and if such a relation can be re-framed to be an indirect 
 experimental evidence towards presence or absence of supersymmetry
 in real world 
 nature). We also like to demonstrate the effect of confinement (as reported earlier for large N non-supersymmetric YM theories via holography) towards  holographic Schwinger particle decay and look
 into exotic results if any. For our purpose 
 the virtual
$q\bar{q}$ pairs are imagined to be endpoint of a string in the boundary. We calculate the rectangular Wilson loop in space-time direction to find out the inter-quark potential. To 
account for an external electric
field we add an extra term. We analytically find out the critical electric field from the same. We also plot figures to illustrate the tunneling phenomenon. Next we move on to finding 
out the critical electric field
from analysis of the DBI action using the fact that the action should be real valued. Then we move on to finding out the circular Wilson loop. It is impossible to do so without any
simplification. We thus expand the
expressions to first order of the Non-SuSy deformation parameter $(u_0)^4$. Doing so we explicitly find out the profile of circular Wilson Loop upto first order of $(u_0)^4$ from
which we find out the pair production rate.
\\
\\
This paper is organized as follows. In section \ref{section2} we recap Non-SuSy D3 branes and their decoupling limit from supergravity. We also show that the Non-SuSy solution goes over to usual AdS when
appropriate limits are taken. In section \ref{section3} we show the derivation of pair production in theory with U(1) gauge field coupled to charged matter. Relevant expression for 
large N gauge theory is also given. In section \ref{section4} we carry 
on Potential Analysis of virtual $q\bar{q}$ pairs from which the critical electric field is derived both by analytical and numerical means. In section \ref{section5} we use the DBI action and find out the critical
electric field using the fact that the action should be real valued. In section \ref{section6} we use perturbative analysis to find out the profile for circular Wilson loop when the string ends at a finite
position ($u_b$). Using this we find the critical electric field and pair production rate and make some comments about the later. We close this paper with conclusions in section \ref{section7}.

\section{Non SuSy Dp Branes and their Decoupling Limit}
\label{section2}

In this section we will take a brief recap of non-supersymmetric Dp brane solutions \cite{Lu:2004ms} and show how to recover the BPS 
Dp brane solutions from them. Then we will state the decoupling limit of Non-SuSy D3 branes by analogy with the BPS case and make sure the the decoupling goes over to the BPS brane decoupling limit when SuSy is
restored\cite{Nayek:2016hsi}. In addition we also show by taking suitable co-ordinate transformation that the decoupled throat geometry is actually identical with two parameter solution obtained
previously by Constable and Myres in which supersymmetry and conformal symmetry are both broken \cite{Constable:1999ch}.
We start with the action for ten dimensional type II supergravity which in addition to the string frame metric $g_{\mu\nu}$, has a dilation $\phi$ field and a $(8-p)$ RR from gauge field $F_{[8-p]}$.
\begin{align}
 \label{sugra}
 S=\frac{1}{16 \pi G_{10}}\int d^{10}x \sqrt{-\text{det}g_{\mu\nu}}\Big[R-\frac{1}{2}\partial_{\mu}\phi \partial^{\mu}\phi-\frac{1}{2(8-p)!}F_{[8-p]}^{2}  \Big]
\end{align}
We will be looking for solutions of the above using the ansatz,
\begin{align}
\label{ansatz}
& ds_{str}^{2}=e^{2A(r)}\Big(-dt^{2}+dx_{1}^{2}+...+dx_{p}^{2}\Big)+e^{2B(r)}\Big(dr^{2}+r^{2}d\Omega_{8-p}^{2}\Big)
\\
& F_{[8-p]}=Q \text{Vol} \big(\Omega_{8-p}\big)
\end{align}
In the above the metric has an ISO(p,1) $\times$ SO(9-p) isometry and represents a magnetically charged p brane in 10 dimensions with magnetic charge $Q$. It can be shown that the above solution conserves super-symmetry 
i.e. saturates the BPS bound if \cite{Duff:1994an}, 
\begin{align}
 \label{susycondition}
 (p+1)B(r)+(7-p)A(r)=0
\end{align}
Solution of equations of \eqref{sugra} compatible with \eqref{ansatz}-\eqref{susycondition} leads to usual BPS p branes. We will be looking for solutions with break 
the condition \eqref{susycondition}, and thus breaks
spacetime supersymmetries. In the rest of the paper we will be concerned with non-supersymmetric D3 brane solution and thus will consider the case where $p=3$.
The non-supersymmetric D3 brane solution is given as,
\begin{align}
 \label{D3}
& ds^{2}=\tilde{F}(\rho)^{-\frac{1}{2}}G(\rho)^{\frac{\delta}{4}}\bigg[-dt^{2}+dx_{1}^{2}+dx_{2}^{2}+dx_{3}^{2}\bigg]+
\tilde{F}(\rho)^{\frac{1}{2}}G(\rho)^{\frac{1+\delta}{4}}\bigg[\frac{d\rho^{2}}{G(\rho)}+\rho^{2}d\Omega_{5}^{2}\bigg]
\nonumber\\
& e^{2\phi}=g_{s}^{2}G(\rho)^{\delta}\;\;\;\;\;\;\;\;\;\;\;\;\;\;\;;\;\;\;\;\;\;\;\;\;\;\;\;\;\;\;F_{[5]}=\frac{1}{\sqrt{2}}(1+\star)\text{ Q Vol}(\Omega_{5})
\end{align}
In the above the functions $\tilde{F}(\rho)$ and $G(\rho)$ are given as,
\begin{align}
\label{functions}
 &\tilde{F}(\rho)=G(\rho)^{\frac{\alpha}{2}}\cosh^{2}\theta-G(\rho)^{-\frac{\beta}{2}}\sinh^{2}\theta \nonumber\\
 &G(\rho)=1+\frac{\rho_{0}^{4}}{\rho^{4}}
\end{align}
It can be shown that the non-SuSy solution \eqref{D3} violates the condition \eqref{susycondition} and thus breaks spacetime supersymmetries.
In the above $e^{2\phi}$ is the effective string coupling constant and the solution is characterized by six parameters
i.e, $\alpha, \beta, \delta, \theta, \rho_{0}, \text{Q}$, of which $\rho_{0}$ has the dimensions of length, Q has 
dimensions of four volume and others are dimensionless. One should further note from \eqref{functions} that the solution given above has a naked singularity at $\rho=0$ and the physical
region is given by $\rho>0$. 
In the context of string theory one hopes that quantum fluctuations modify the behavior of the solution near the singularity point. As $e^{2\phi}$ is the effective string coupling, for the 
supergravity description to remain 
valid one needs the \textit{parameter $\delta$ to be less or equal to zero} so as to make the string coupling small. The parameters of the solutions are not all independent but satisfy some consistency relations like,
\begin{align}
 \label{constraints}
\alpha=\beta\;\;\;\;\;\;\;\;\;,\;\;\;\;\;\;\;\;\;\text{Q}=2\alpha\rho_{0}^{4}\sinh 2\theta\;\;\;\;\;\;\;\;\;,\;\;\;\;\;\;\;\;\;\alpha^{2}+\delta^{2}=\frac{5}{2}
\end{align}
In arbitrary dimensions the solutions and the constraints are a bit complicated and is given in \cite{Nayek:2015tta}. Just like the BPS D3 brane solution, the non-SuSy solution too is asymptotically
flat. One can recover the
BPS solution from the 
non-SuSy solution given above by considering the limits $\rho_{0}\rightarrow 0$ and $\theta \rightarrow \infty$
keeping $\frac{\alpha}{2}\rho_{0}^{4}(\cosh^{2}\theta+\sinh^{2}\theta)\rightarrow R^{4}=$fixed. Under this
scaling one has $G(\rho)\rightarrow 1$ and $\tilde{F}(\rho)\rightarrow 1+\frac{R^{4}}{\rho^{4}}$ and $\text{Q}\rightarrow 4R^{4}$ under which the standard BPS solution is regained.
\\
\\
The decoupling limit is a low energy limit in which interactions between the bulk theory and theory living on the brane vanishes. To work out the decoupling limit and henceforth the 
throat geometry one needs to make a change of variables in analogy with the BPS D3 brane.
\begin{align}
 \label{decoupling}
 \rho=\alpha^{\prime}u\;\;\;\;\;\;,\;\;\;\;\;\;\rho_{0}=\alpha^{\prime}u_{0}\;\;\;\;\;\;,\;\;\;\;
 \;\;\alpha \cosh^{2}\theta=\frac{\lambda}{{\alpha{\prime}}^{2}u_{0}^{4}}\;\;\;\;\;\;,\;\;\;\;\;\;\alpha^{\prime}\rightarrow 0
\end{align}
In the above $u$ and $u_{0}$ have the dimensions of energy and are kept fixed. From \eqref{constraints} and \eqref{decoupling} it can be shown that $\frac{Q}{{\alpha^{\prime}}^{2}}\gg 1$ implying
that the curvature of spacetime 
in string units must be very small for the supergravity description to be valid.
A justification of the above decoupling limit is 
given explicitly in \cite{Nayek:2016hsi} and \cite{Nayek:2015tta}. Under the above said limit,
\begin{align}
 &G(\rho)\rightarrow G(u)= 1+\frac{u_{0}^{4}}{u^{4}}=\text{fixed} \\
 &\tilde{F}(\rho)\rightarrow \tilde{F}(u)=\frac{\lambda}{{\alpha^{\prime}}^{2}}F(u)
 \end{align}
 In the above $F(u)=\frac{1}{\alpha u_{0}^{4}}\big(G(u)^{\frac{\alpha}{2}}-G(u)^{-\frac{\alpha}{2}}\big)$ and the non-SuSy D3 brane throat geometry in the decoupling limit mentioned above becomes
 \begin{align}
  \label{throat}
  &ds^{2}=\alpha^{\prime}\sqrt{\lambda}\Big[F(u)^{-\frac{1}{2}}G(u)^{\frac{\delta}{4}}\eta_{\mu \nu}dx^{\mu}dx^{\nu}+F(u)^{\frac{1}{2}}G(u)^{\frac{1+\delta}{4}}
  \big(\frac{du^{2}}{G(u)}+u^{2}d\Omega_{5}^{2}\big)\Big]\nonumber\\
  &e^{2\phi}=g_{s}^{2}G(u)^{\delta}
 \end{align}
 In the above the spacetime co-ordinates has been rescaled as $(t,x^{i})\rightarrow \sqrt{\lambda}(t,x^{i})$ where $\lambda$ is the 't hooft coupling. In the 
 limit $u_{0}\rightarrow 0$ one has $G(u)\rightarrow 1$ and 
 $F(u)=\frac{1}{\alpha u_{0}^{4}}\big[\frac{\alpha u_{0}^{4}}{u^{4}}+\mathcal{O}(\frac{u_{0}^{8}}{u^{8}})\big]\approx u^{4}$. In this limit the non-SuSy throat 
 geometry \eqref{throat} goes over to the known 
 AdS$_{5} \times$S$^{5}$, and the effective string coupling becomes constant. To check the relation of solution \eqref{throat} with that of the previously known
 one by Constable and Myres \cite{Constable:1999ch} which was 
 conjectured to be dual to some non-supersymmetric field theory, one has to re-write the solution in the Einstein frame,
 \begin{align}
 \label{Einstein}
  &ds^{2}_{E}=\alpha^{\prime}\sqrt{\lambda}\Big[H(u)^{-\frac{1}{2}}G(u)^{\frac{\alpha}{4}}\eta_{\mu\nu}dx^{\mu}dx^{\nu}+H(u)^{\frac{1}{2}}G(u)^{\frac{1-\alpha}{4}}\Big(\frac{du^{2}}{G(u)}+u^{2}d\Omega_{5}^{2}\Big)\Big]
  \nonumber\\
  &e^{2\phi}=g_{s}^{2}G(u)^{\delta}
 \end{align}
 In the above the function $H(u)$ is defined by $  H(u)=G(u)^{\frac{\alpha}{2}}F(u)=G(u)^{\alpha}-1$.  Now one has to make a co-ordinate transformation like $u=r\Big(1+\frac{\omega^{4}}{r^4}\Big)^{-\frac{1}{4}}$ 
 where $\omega^4=\frac{u_{0}^4}{4}$.
 Under this transformation, $G(u)\rightarrow \Big(1+2\frac{\omega^4}{r^4}\Big)^2$ and $H(u)\rightarrow \Big(1+2\frac{\omega^4}{r^4}\Big)^{2\alpha}-1$. From these
 relations and \eqref{Einstein} one can exactly produce
 the two parameter family of solutions as found in \cite{Constable:1999ch} in which both supersymmetry and conformal symmetry is broken. The solution in \cite{Constable:1999ch}
 also exhibits QCD like behavior like 
 running gauge coupling and confinement in the infrared. The geometry \eqref{Einstein} exhibits a naked singularity at $u=0$, and thus should be corrected by stringy corrections
 which should become dominant at low length
 scales. Moreover the proper distance (spatial) from the exterior (say $u=u_b$) to the interior is finite (which says that stringy corrections are a must). In holography the proper
 distance is identified with mass of the
 string hanging from the boundary to the interior \cite{Kinar:1998vq}. To find the same we have to choose a gauge of the form : $x_{0}=t$, $u=s$, $\text{all others}=\text{constant}$. With 
 this gauge the mass is given by
\begin{align}
 \label{massactual}
 m=\frac{\sqrt{\lambda}}{2\pi}\int_{0}^{u_b}\;du\sqrt{\bigg(1+\frac{u_0^4}{u^4}\bigg)^\frac{2\delta-3}{4}}=\text{finite and positive for all allowed values of}\;\delta.
\end{align}
The integral can indeed be done in closed form. However the result is very complicated (hypergeometric functions involved), and it is very difficult to invert $u_b$ in terms of $m$. Thus we express our results in 
this work with formula for mass($m_0$) of $\mathcal{N}=4$ SYM.
\begin{align}
 \label{mass}
 m_0=\frac{\sqrt{\lambda}}{2\pi}u_b
\end{align}

\section{Pair Production in presence of External Fields}
\label{section3}
In this section we will revisit the concept of pair production in presence of external electric fields i.e the "Schwinger Effect". We will demonstrate the effect using euclidean version of
the electromagnetic action \cite{Affleck:1981bma} and generalize to large N gauge theories. The euclideanized version of U(1) gauge theory coupled to a massive complex scalar field is given by
\begin{align}
 S=\int d^4x\Big[\frac{1}{4}F_{\mu\nu}^2+|(\partial_\mu+ieA_\mu+iea^{ex}_\mu)\phi|^2+m^2|\phi|^2\Big]
\end{align}
In the above $A_\mu$ refers to the dynamical U(1) gauge field and $a_{\mu}^{ex}$ refers to the external value of (constant) electromagnetic field. The pair production rate, $\Gamma$ can
be written as \cite{Itzykson:1980rh}
\begin{align}
V\Gamma &=-2\;\text{Im ln}\int \mathcal{D}A\mathcal{D}\phi\;e^{-S} \nonumber\\
&=-2\;\text{Im ln}\int \mathcal{D}A e^{-S_{eff}}
\end{align}
Where, $S_{eff}=\frac{1}{4}\int d^{4}xF_{\mu\nu}^{2}+ \text{tr ln}\big[-\big(\partial_{\mu}+ieA_{\mu}+iea^{ex}_{\mu}\big)^{2}+m^2\big]$. For leading order calculations one can ignore the coupling of the 
dynamical gauge field with the scalar field. Thus the expression above reduces to
\begin{align}
 V\Gamma=-2\;\text{Im tr ln}\Big[-(\partial_\mu+iea^{ex}_{\mu})^2+m^2\Big]
\end{align}
Using the relation, $\text{tr ln}(A)=-\int_{0}^{\infty}\;\frac{dT}{T}\;\text{tr}e^{-AT}$ and evaluating the trace in position basis, one can rewrite the above expression to
\begin{align}
 V\Gamma=\text{Im}\int_{0}^{\infty}\;\frac{dT}{T}\;e^{-\frac{m^2T}{2}}\int d^4x\;\langle x| \exp\bigg[-T\Big\{-(\partial_{\mu}+iea^{ex}_{\mu})^2\Big\}\bigg]|x\rangle
\end{align}
Note that the integrand under $d^4x$ is synonymous to the path integral of a non-relativistic particle under the influence of the Hamiltonian $H=\frac{1}{2}\Big[P_\mu+ea^{ex}_\mu\Big]^2$. Using
quantum mechanical path
integral representation \cite{Schubert:2001he}, one can write
\begin{align}
V\Gamma=&\text{Im}\int_{0}^{\infty}\;\frac{dT}{T}\;e^{-\frac{m^2T}{2}}\int_{x(0)=x(T)}\mathcal{D}x \exp\Big[-\frac{1}{2}\int_{0}^{T}d\tau\dot{x}^{2}+ie\oint a^{ex}_\mu dx_\mu\Big]\nonumber\\
=&\text{Im}\int_{0}^{\infty}\;\frac{dT}{T}\int_{x(0)=x(1)}\mathcal{D}x \exp\Big[-\frac{1}{2T}\int_{0}^{1}d\tau\dot{x}^{2}-\frac{m^2T}{2}+ie\oint a^{ex}_\mu dx_\mu\Big]
\end{align}
 Where in the last line we have rescaled $\tau \rightarrow \frac{1}{T}\tau$. We assume $m^2\int_{0}^{1}d\tau \dot{x}^{2}\gg1$ (a condition signifying heavy mass) and note that the integration 
 over $T$ has the form
 of a modified Bessel function $K_{0}(x)=\int_{0}^{\infty} \frac{dt}{t} \exp \Big(-t-\frac{x^{2}}{4t}\Big)$ with the asymptotic behavior, $K_{0}(x)\simeq \sqrt{\frac{\pi}{2x}}e^{-x}$, for 
 large $x$. Thus the above integral becomes
 \begin{align}
  V\Gamma=\text{Im} \int \mathcal{D}x\exp \big[-S_p\big]\frac{1}{m}\sqrt{\frac{2\pi}{T_{0}}}
 \end{align}
 In the above , $T_{0}=\frac{1}{m}\sqrt{\int d\tau \dot{x}^{2}}$ and $S_p=m\sqrt{\int d\tau \dot{x}^{2}}-ie\oint a^{ex}_\mu dx_\mu$ and $a_{1}^{ex}=-iEx_{0}$ (signifying
 constant electric field of value E in $x_1$ direction, iota comes in due to euclidean signature). We like to evaluate the above integral by the method of steepest descent. The argument within the exponential is 
 the action for a relativistic particle executing a periodic motion under influence of $a_{\mu}^{ex}$. The equation of motion for it is given by
 \begin{align}
  \frac{1}{\sqrt{\int \dot{x}^{2}}}m\ddot{x}_{\mu}=eF^{ex}_{\mu\nu}\dot{x}_{\nu}
 \end{align}
Keeping in mind the periodic boundary conditions $x_{\mu}(0)=x_{\mu}(1)$, $F^{ex}_{01}=E$  one has the following classical solution
\begin{align}
 x_{\mu}^{cl}=R(0,0,\cos 2\pi \tau, \sin 2 \pi \tau)\,\,\,\,\,,\,\,\,\,\,\text{R}=\frac{m}{eE}\,\,\,\,\,,\,\,\,\,\,S_p^{cl}=\frac{\pi m^2}{eE}
\end{align}
Using the above values one has $\frac{1}{m}\sqrt{\frac{2\pi}{T_{0}}}=\frac{\sqrt{eE}}{m}$. Thus decay rate can be approximated as
\begin{align}
 V\Gamma\approx\frac{\sqrt{eE}}{m}e^{-\frac{\pi m^2}{eE}}
\end{align}
Ideally one should go around calculating the one loop prefactor and complete the steepest descent process \cite{Affleck:1981bma},\cite{Dunne:2006st}, the 
calculation of which is indeed complicated. The modified prefactor is given by $\frac{(eE)^{2}}{(2\pi)^3}$. Thus we see that the pair production rate does to zero if the external electric field is switched off.
In arbitrary coupling one can no longer neglect the effect of the dynamical fields and one has to include contribution from Wilson loops
\begin{align}
V\Gamma=-2\text{Im}\int_{0}^{\infty}\frac{dT}{T}e^{-\frac{m^2T}{2}}\int \mathcal{D}x \exp\Big[-\frac{1}{2T}\int_{0}^{1}d\tau\dot{x}^{2}+ie\oint a^{ex}_\mu dx_\mu\Big]\langle \exp \Big(ie\oint A_\mu dx_\mu\Big)\rangle
\end{align}
The pair production rate gets modified to \cite{Affleck:1981bma}\cite{Kawai:2015mha}
\begin{align}
\label{fluc}
 \Gamma=\frac{(eE)^{2}}{(2\pi)^3}\sum_{n=1}^{\infty}\frac{(-1)^{n+1}}{n^2}\exp \Big(-n\big(\frac{\pi m^2}{eE}-\frac{e^2}{4}\big)\Big)
 \end{align}
 From the above one can work out that the pair production rate is not exponentially suppressed once the value of electric field exceeds the so called critical value $E_{c}=\frac{4\pi m^2}{e^3}$, beyond which the 
 vacuum becomes unstable.
 \\
 \\
To implement this argument for AdS/CFT like theories one faces a number of problems. Firstly the field theory in those circumstances is a conformal one and one cannot get a mass term a priori. Moreover in
the dual gauge theory matter fields exists in the adjoint representation of SU(N) gauge group. To evade these issues one uses the Higgs mechanism to break the symmetry group from SU(N+1)$\rightarrow$SU(N)
$\otimes$U(1). Because of this splitting one has $5$ massive W bosons transforming in fundamental representation of SU(N) and interacting with the background Yang Mills theory. Now the pair production rate
in presence
on an external electric field is given by \cite{Kawai:2015mha} \cite{Chowdhury:2019mqi}
\begin{align}
\label{exact}
 \Gamma \sim& -5\text{N}\;\int \mathcal{D}x \;\text{exp}\Big(-m\int_{0}^{1}d \tau\;\sqrt{\dot{x}^{2}}+
 i\int_{0}^{1}d \tau \;a^{(E)}_{\mu}\dot{x}_{\mu}\Big)\langle W[x] \rangle
\end{align}
Where W[x] is the SU(N) Wilson loop and can be calculated by holographic means.

\section{Pair Production in Non-supersymmetric Theories via Holography}
\label{section4}

The ideal way to argue Schwinger effect\cite{Semenoff:2011ng}\cite{Bolognesi:2012gr} is to calculate the expectation value of circular Wilson loops and relate it to the decay rate. 
However one can alternatively view the vacuum to be made of virtual $q\bar{q}$
pairs in presence of an attractive potential and study the influence of an external electric field \cite{Sato:2013iua}. This basically amounts to calculating the inter-quark potential which one does by considering the
rectangular Wilson loop. In doing so one has to make some additional approximations. One considers that the time scale associated with the Wilson loop is much lesser than the length scale. Intuitively, one thinks that 
the quark anti-quark pairs are separated in the far past and unite in the far future. In holography the Wilson loop is given by following formula \cite{Maldacena:1998im} \cite{Drukker:1999zq}
\begin{align}
\label{ideal}
\langle{W[\mathcal{C}]}\rangle=\frac{1}{\text{Vol}}\int_{\partial X=\mathcal{C}} \mathcal{D}X \mathcal{D}h_{ab}\;e^{-S[X,h]}
\end{align}
$S[X,h]$ is the Wick rotated action of the fundamental string \cite{Blumenhagen:2013fgp} with endpoints ending at contour $\mathcal{C}$ situated on the probe brane. In the classical
limit ($\alpha^{\prime}\rightarrow 0$)
the extremal value of the string action dominates and thus the Wilson loop is the extremal area of string world-sheet ending on the contour. To study the rectangular Wilson loop we take 
the quark anti-quark dipole to
be aligned in the $x_{3}$ direction. The string action whose on-shell value we are interested
with is the Nambu-Goto action $\mathcal{S}_{NG}=\frac{1}{2 \pi \alpha^{\prime}}\int dt ds \sqrt{det\;G^{(in)}_{ab}}$
with $G^{(in)}_{ab} \equiv G_{\mu\nu} \frac{\partial x^{\mu}}{\partial s^{a}} \frac{\partial x^{\nu}}{\partial s^{b}}$ which has two diffeomorphism symmetries. We
exploit those to choose the following gauge
\begin{align}
 \label{staticgauge}
 x^{0}(s,t)=t\;\;;\;\;x^{3}(s,t)=s\;\;;\;\;u(s,t)=u(s)\;\;;\;\;x^{1,2}=0\;\;;\;\;\Theta^{i}(s,t)=\text{constant}
\end{align}
\begin{figure}[tbp]
\centering 
\includegraphics[width=1.00\textwidth,height=0.60\textwidth,origin=c,angle=0]{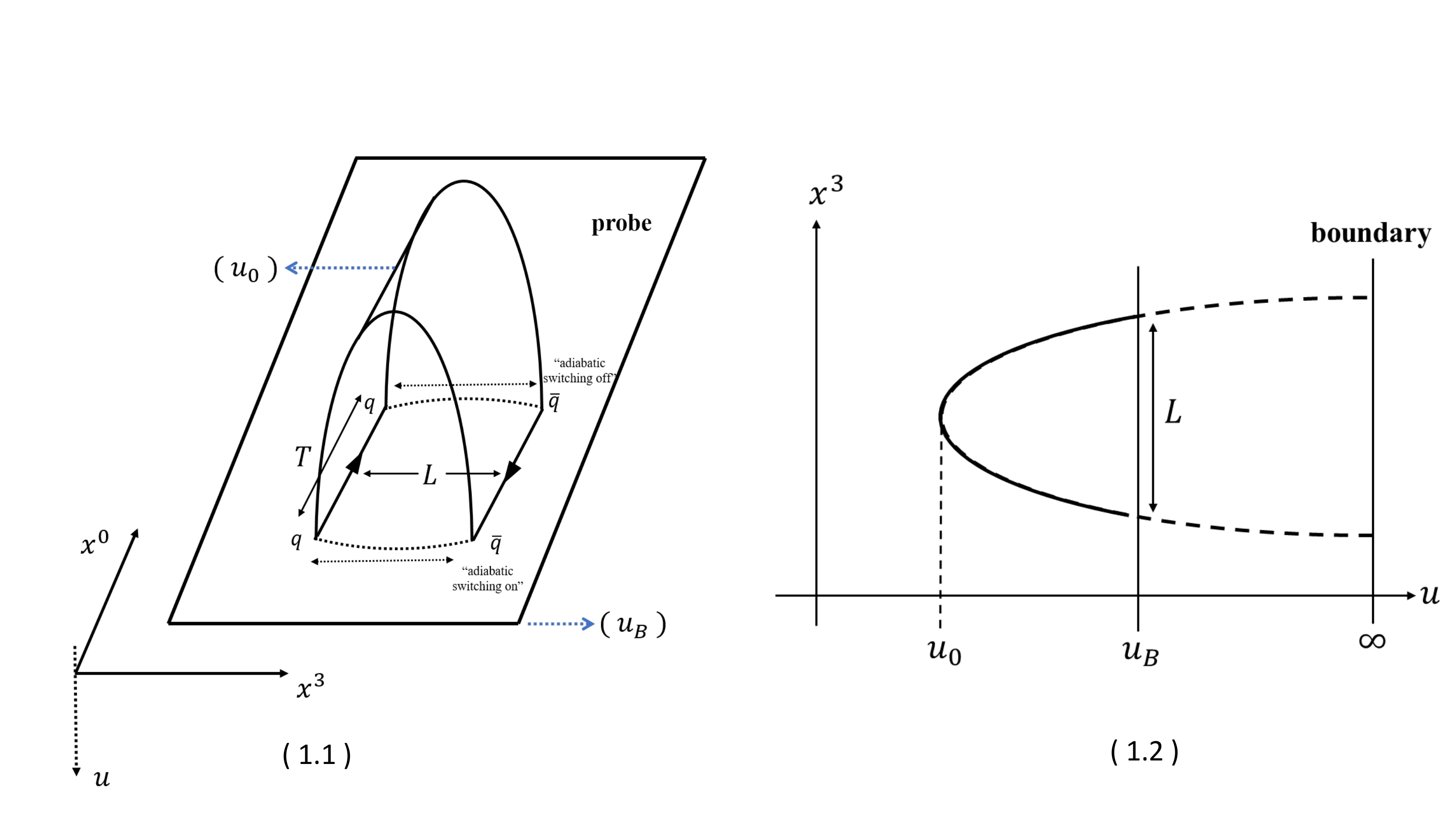}
\caption{This figure illustrates the setup used. The probe brane is placed at a finite position ($u_{b}$) on the holographic direction as in (1.2). On the probe brane
 the placement of the Wilson loop is shown in (1.1), arrows indicating the contour of the loop (not the propagation of the string). For adiabatic interactions one can neglect the effects of the
 dotted lines and the string profile becomes static. }
\label{fig1}
\end{figure}
For present purposes $x_{3} \equiv s $ is assumed to to range between $[-L,L]$ and temporal direction $x_{0} \equiv t $ is 
ranged between $[-\mathcal{T},\mathcal{T}]$ with the assumption that $\mathcal{T}\gg L$. $2L$ indicates the inter-quark separation on the probe brane with the boundary 
condition $u(\pm L)=u_{b}$ where $u_{b}$ indicates the position of the probe brane along the holographic direction, see Figure\ref{fig1}. Finally another word about the configuration, it is possible to consider 
the $q\bar{q}$ pairs at a velocity in the $x_{2}$ direction. However in the present case where the virtual particles in vacuum are modeled as $q\bar{q}$ dipoles, such a configuration seems hardly sensible. 
The induced metric as per the above gauge choice reads \eqref{throat}\eqref{staticgauge}, 
\begin{align}
 \frac{1}{\alpha^{\prime}\sqrt{\lambda}}G^{(in)}_{ab} ds^{a}ds^{b}=-F(u(s))^{-\frac{1}{2}}&G(u(s))^{\frac{\delta}{4}}dt^{2}+ds^{2}\bigg[F(u(s))^{-\frac{1}{2}}G(u(s))^{\frac{\delta}{4}}\nonumber\\
 +&F(u(s))^{\frac{1}{2}}G(u(s))^{\frac{1+\delta}{4}}\frac{1}{G(u(s))}\left(\frac{du}{ds}\right)^{2}
\bigg]
\end{align}
From the above we have the determinant of the induced metric to be 
\begin{align}
\label{rinduced}
 -\text{det}G^{(in)}_{ab}=(\alpha^{\prime}\sqrt{\lambda})^{2}\bigg[G(u(s))^{\frac{2\delta-3}{4}}\bigg(\left(\frac{du}{ds}\right)^{2}+G(u(s))^{\frac{3}{4}}F(u(s))^{-1}\bigg)\bigg]
\end{align}
It is not possible to carry on analysis without some simplification. We therefore assume that $\left(\frac{u_{0}}{u}\right)^{4}\ll 1$ and with the mentioned, simplify the area i.e. on shell Nambu-Goto action to
\begin{align}
\label{ng}
 \mathcal{S}_{ng}&=\frac{1}{2\pi\alpha^{\prime}}\int_{-\mathcal{T}/2}^{\mathcal{T}/2}dt\int_{-L}^{L}ds\;\sqrt{-\text{det}G^{(in)}_{ab}}\nonumber\\
 &=\frac{\sqrt{\lambda}}{2\pi}\mathcal{T}\int_{-L}^{L}ds\;\sqrt{\left(\frac{du}{ds}\right)^{2}\bigg(1+A\frac{u_{0}^{4}}{u^{4}}\bigg)+u^{4}\bigg(1+B\frac{u_{0}^{4}}{u^{4}}\bigg)}
\end{align}
Wherein
\begin{align}
 B=\frac{\delta+1}{2}\;\;\;\;\;;\;\;\;\;\;A=\frac{2\delta-3}{4}\;\;\;\;\;;\;\;\;\;\;A+\frac{5}{4}=B
 \end{align}
Crudely speaking, this can be seen as treating the Non-Susy theory as perturbation over the $\mathcal{N}$=4 supersymmetric Yang Mills. Since the expression \eqref{ng} doesn't explicitly
depend on the parameter $s$, the corresponding "Hamiltonian", $Q$ is conserved.
\begin{align}
\label{charge}
 Q=-\frac{du}{ds}\frac{dL_{ng}}{d\left(\frac{du}{ds}\right)}+L_{ng}=\frac{u^{4}+Bu_{0}^{4}}{\sqrt{\left(\frac{du}{ds}\right)^{2}\bigg(1+A\frac{u_{0}^{4}}{u^{4}}\bigg)+u^{4}\bigg(1+B\frac{u_{0}^{4}}{u^{4}}\bigg)}}
\end{align}
A indicated in \cite{Rey:1998ik} the fundamental string is assumed to carry charges at two of its endpoints and is otherwise symmetric about its origin. From the above expression we see that
$\frac{du}{ds}$ has both positive and negative sign. Appealing to its symmetric nature there exists a point, namely turning point ( with string parameter $s_{t}$ ) such that
\begin{align}
\bigg(\frac{du}{ds}\bigg)(u_{t})=0
\end{align}
Using the above expression in \eqref{charge} the value of the conserved Hamiltonian is found in terms of the turning point
\begin{align}
 Q=\sqrt{u_{t}^{4}+Bu_{0}^{4}}
\end{align}
Putting the above value in \eqref{charge} we get
\begin{align}
 \label{slope}
 \frac{du}{ds}=u^2\sqrt{\frac{(u^4-u_{t}^4)(u^{4}+Bu_{0}^4)}{(u_{t}^{4}+Bu_{0}^4)(u^4+Au_{0}^4)}}
\end{align}
The length of the (virtual) dipole can be calculated to be (see Figure \ref{fig1})
\begin{align}
 \label{length}
 L=\int_{-L/2}^{L/2}\;dx_{3}=\sqrt{u_t^{4}+Bu_{0}^{4}}\int_{u_t}^{u_b}du\frac{\sqrt{u^4+Au_0^4}}{u^2\sqrt{(u^4-u_t^4)(u^4+Bu_0^4)}}
\end{align}

\begin{figure}[tbp]
\centering 
\includegraphics[width=0.6\textwidth,height=0.50\textwidth,origin=c,angle=0]{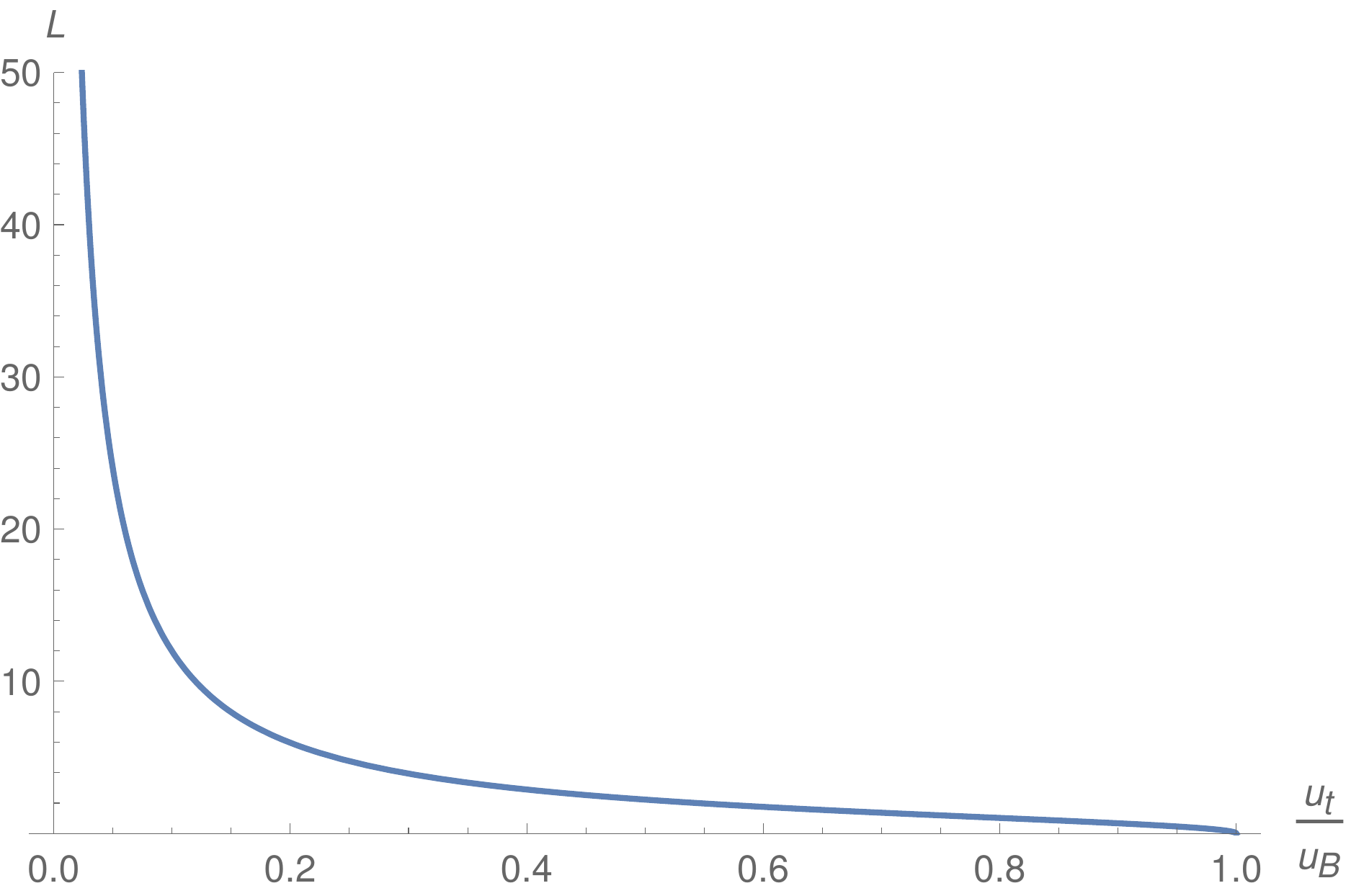}
\caption{This is the graph of L v/s $u_t$. Note that the function is an isomorphism. The values used are $\delta=-0.75$ and $\frac{u_0}{u_b}=0.01$}
\label{fig7}
\end{figure}

\begin{figure}[tbp]
\centering 
\includegraphics[width=0.5\textwidth,height=0.60\textwidth,origin=c,angle=0]{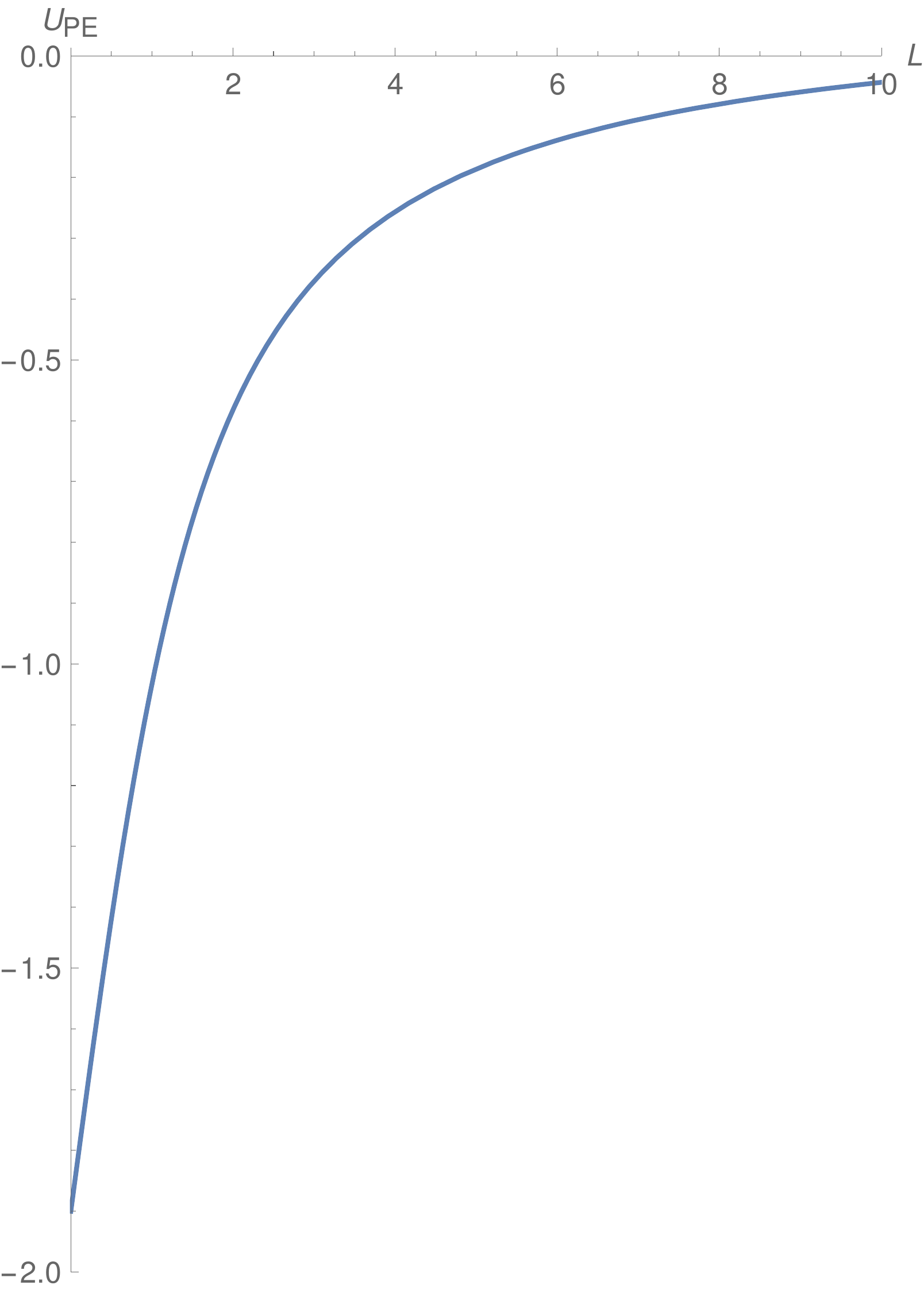}
\caption{This is the graph of $U_{PE}$ v/s L. The rest mass has been duly subtracted. Note that for small values of L the graph is approximately linear and for large L coulombic
behavior is mimicked. Deviation from 
usual coulombic behavior is evident.The values used are $\delta=-0.75$, $\frac{u_0}{u_b}=0.01$ and $\lambda=4\pi^{2}$. }
\label{fig8}
\end{figure}

\begin{figure}[tbp]
\centering 
\includegraphics[width=0.6\textwidth,height=0.75\textwidth,origin=c,angle=0]{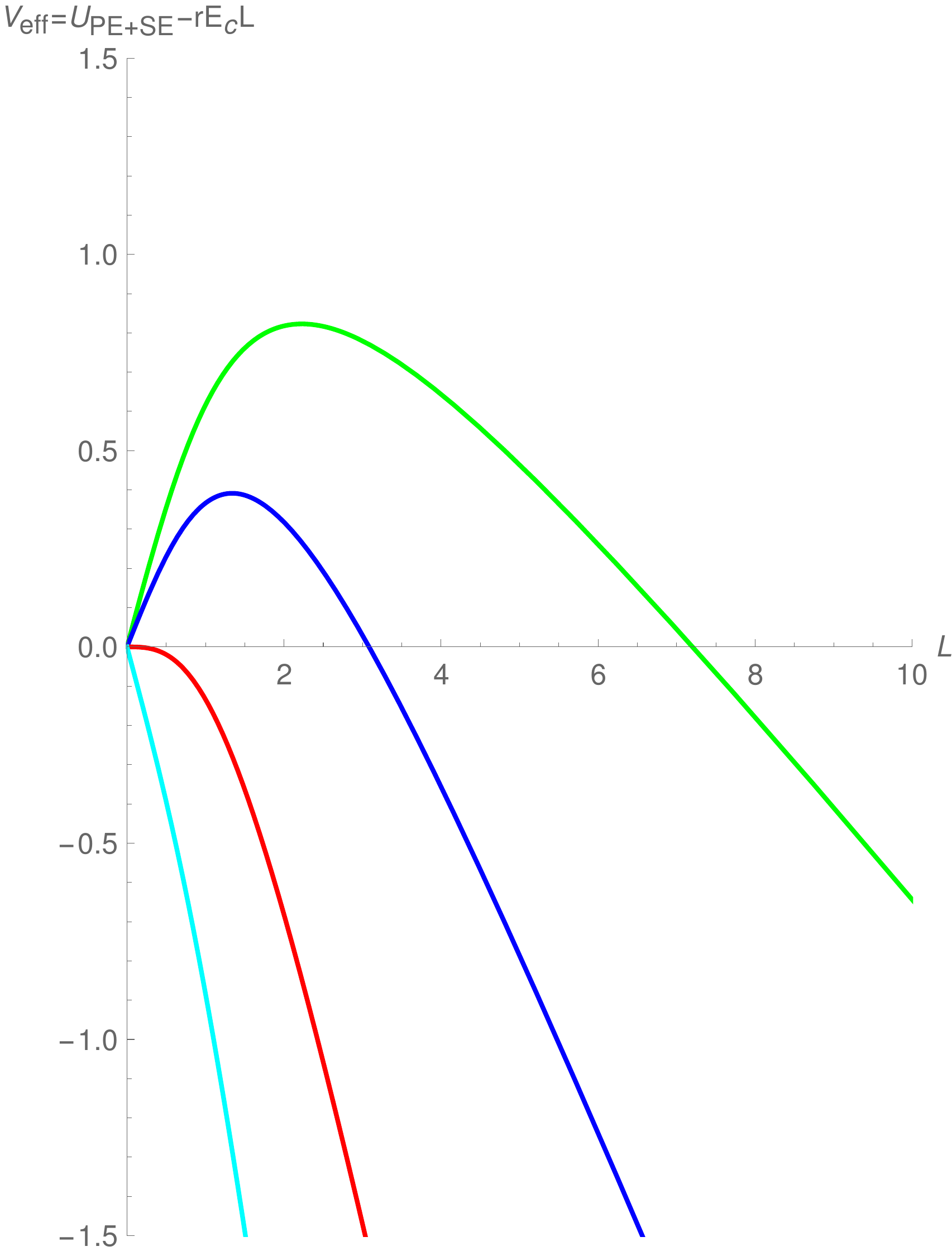}
\caption{The plot indicates the effective potential (in presence of external electric field) v/s the inter-quark separation. Imagine this to be then potential through which q$\bar{q}$ tunnels out. 
The \textcolor{green}{green} line indicates \textcolor{green}{r=0.25}, \textcolor{blue}{blue} line for \textcolor{blue}{r=0.75}. The parameter
r is the ratio of the applied
field to it's threshold value. The 
\textcolor{red}{red} line which exhibits the threshold behavior i.e. no potential barrier stands for \textcolor{red}{r=1.0} and \textcolor{cyan}{cyan} for \textcolor{cyan}{r=1.75} shows
catastrophic decay of vacuum. Note that at the threshold/ critical value, the slope of the potential vanishes at $L=0$ and is negative for nonzero value of L which is precisely 
the conditions we have used to analytically find out the value of $E_c$. Also note that none of the above plots exhibit confining behavior as earlier reported in literature! The values 
used are $\delta=-0.75$, $\frac{u_0}{u_b}=0.01$ and $\lambda=4\pi^{2}$. }
\label{fig9}
\end{figure}

From, \eqref{slope} and \eqref{ng} we can find the on-shell value of inter-quark potential ,
\begin{align}
 \label{potential}
 \mathcal{U}_{PE+SE}=\frac{\mathcal{S}_{ng}}{\mathcal{T}}=\frac{\sqrt{\lambda}}{2\pi}\int_{u_t}^{u_b}du\;\sqrt{(u^4+Au_{0}^4)(u^{4}+Bu_{0}^4)}\frac{1}{u^2\sqrt{u^4-u_t^4}}
\end{align}
Notice from \eqref{length} that when $u_t \rightarrow u_b$, the value of the inter-quark separation becomes small. But as said earlier we are in an approximation where $\left(\frac{u_{0}}{u}\right)^{4}\ll 1$. 
Thus the
calculations in this section are trustable for large inter-quark separation. Now the expression in \eqref{potential} (see Figure \ref{fig8} for the plot) doesn't take the presence of an external electric field into account. Thus we define
an effective potential as
\begin{align}
 \label{effectivepotential}
 \mathcal{V}_{eff}=\mathcal{U}_{PE+SE}-E.L=(1-r)E_c.L+G(u_t(L))
\end{align}
In the above we have assumed the presence on an critical electric field $E_c$, above which the effective inter-quark force becomes repulsive for all values of the inter-quark separation. 
The quantity $G(u_t)$ is 
\begin{align}
 \label{G}
 G(u_t)=\mathcal{U}_{PE+SE}-E_c L=\int _{u_t}^{u_b}du\frac{\sqrt{u^4+Au_0^4}}{u^2\sqrt{u^4-u_t^4}}\Big[\frac{\sqrt{\lambda}}{2\pi}\sqrt{u^4+Bu_0^4}-E_c\frac{\sqrt{u_{t}^4+Bu_0^4}}{\sqrt{u^4+Bu_0^4}}\Big]
\end{align}
The parameter r is the ratio of applied electric field to its critical value. The slope of the effective potential is given as
\begin{align}
 \label{force}
 \frac{d\mathcal{V}_{eff}}{dL}=(1-r)E_c+\frac{du_t}{dL}\frac{dG(u_t)}{du_t}
\end{align}
We now proceed to find the value of the critical electric field. Note, at $u_t=u_b$ the inter quark separation \eqref{length} and the inter-quark potential \eqref{potential} vanishes, 
see Figure \ref{fig7}. 
At the critical value of the electric field $r=1$, the $1^{st}$ term of \eqref{force} ceases to contribute, and the behavior of the inter-quark force will be completely governed by the 
second term of
\eqref{force}. Criticality demands that the potential ceases to put up a tunneling barrier for all values of inter-quark separation, see red line in Figure \ref{fig9}. Given that $G(u_t(L))$ 
vanishes at $L=0$  we need
to show that $G(u_t(L))$ is a monotonically decreasing function with respect to L whose slope vanishes at $L=0$. (This is because critical
electric field is the least one for which pair production happens spontaneously).
From \eqref{length} we have
\begin{align}
\label{dlength}
 \frac{dL}{du_{t}}=-\frac{\sqrt{u_t^4+Au_0^4}}{u_t^2\sqrt{\big(u_t^4+\epsilon\big)^4-u_t^4}}+2\int_{u_t+\epsilon}^{u_b}du\;\frac{u_t^3}{u^2}
 \frac{\sqrt{\big(u^4+Au_0^4\big)\big(u^4+Bu_0^4\big)}}{\sqrt{\big(u^4-u_t^4\big)^3\big(u_t^4+Bu_0^4\big)}}
\end{align}
Similarly we have from \eqref{G}
\begin{align}
 \label{dG}
 \frac{dG}{du_t}(u_t)=-&\frac{\sqrt{u_t^4+Au_0^4}}{u_t^2\sqrt{(u_t+\epsilon)^4-u_t^4}}\Big[\frac{\sqrt{\lambda}}{2\pi}\sqrt{u_t^4+Bu_0^4}-E_c\Big]\nonumber\\
 &+2u_t^3 \int_{u_t+\epsilon}^{u_b}du\;\frac{\sqrt{(u^4+Au_0^4)(u^4+Bu_0^4)}}{u^2\big(\sqrt{u^4-u_t^4}\big)^3}\Big[\frac{\sqrt{\lambda}}{2\pi}-\frac{E_c}{\sqrt{u_t^4+Bu_0^4}}\Big]\nonumber\\
 =&\Big[\frac{\sqrt{\lambda}}{2\pi}\sqrt{u_t^4+Bu_0^4}-E_c\Big]\frac{dL}{du_t}
\end{align}
Thus we get
\begin{align}
\label{finalforce}
\frac{d\mathcal{V}_{eff}}{dL}=(1-r)E_c+  \Big[\frac{\sqrt{\lambda}}{2\pi}\sqrt{u_t^4+Bu_0^4}-E_c\Big]           
\end{align}
At threshold condition the slope of the potential should be zero at when inter-quark separation vanishes i.e. $u_t=u_b$. Implementing the same in \eqref{finalforce} we get
\begin{align}
 \label{pcritical}
 E_{c}=&\frac{\sqrt{\lambda}}{2\pi}u_{b}^{2}\sqrt{1+\frac{\delta+1}{2}\frac{u_{0}^{4}}{u_{b}^{4}}}\nonumber\\
 =& \frac{2\pi}{\sqrt{\lambda}}m_0^2\sqrt{1+\frac{\lambda^2}{32\pi^4}\big(\delta+1\big)\frac{u_0^4}{m_0^4}}
\end{align}
We thus have ,
\begin{align}
 \label{finalfinalforce}
 \frac{d\mathcal{V}_{eff}}{dL}=(1-r)E_c+  \Big[\frac{\sqrt{\lambda}}{2\pi}\sqrt{u_t^4+Bu_0^4}-\frac{\sqrt{\lambda}}{2\pi}\sqrt{u_b^4+Bu_0^4}\Big]  
\end{align}
From Figure \ref{fig7} we see that L increases as $u_t$ decreases using which we can say from \eqref{finalfinalforce} that $\frac{d\mathcal{V}_{eff}}{dL}$ is a monotonically 
decreasing function of L at $r=1$. It can be easily
understood that from $r>1$ the effective potential in totally repulsive. Thus we establish the existence of
a critical electric field with value given by \eqref{pcritical}.
We see that as $\delta$ switches over $-1$, the critical electric field increases and decreases respectively compared to the supersymmetric value. Not even that, just at $\delta=-1$, the critical field has 
the same value
as that of the supersymmetric theory. Will this kind of behavior remain when one considers higher orders? How much of the calculation in this section
should be trusted for small values of inter-quark separation? The answer to this question will be found in the next section.
\\
\\
It so happens that analytical solutions to \eqref{length},\eqref{potential},\eqref{effectivepotential} cannot be found out in a closed from via Mathematica. Thus we resort to
numerical methods. Some plots to illustrate 
the situation are given.

\section{DBI Analysis of Critical Electric Field}
\label{section5}
In this section we look to find out the critical electric field from analysis of the DBI action of the probe brane in presence of an external electric field. In due course we  will also answer the 
question raised in section \ref{section4}. 

As earlier we imagine the probe brane situated at $u=u_{b}$  (see Figure \ref{fig1}) in the holographic dual with an electric field switched on at the brane position. The DBI action is given as 
\begin{align}
 \mathcal{S}_{\text{DBI}}=\frac{1}{(2\pi)^{3}g_{s}\alpha^{\prime}}\int_{u=u_{b}} d^{4}x \sqrt{-\text{det}(P[g]_{\mu\nu}+B_{\mu\nu}+2\pi \alpha^{\prime}F_{\mu\nu} )}
\end{align}
In the above$P[g]_{\mu\nu}$ is the pullback of the curved metric on the probe brane, $B_{\mu\nu}$ is the NS 2-form which is zero in the present case. $F_{\mu\nu}$ is the Faraday tensor which we set
to the value $F_{03}=E$, to indicate the presence of an external electric field. Evaluating the above from \eqref{throat} we have

\begin{align}
&P[g]_{\mu\nu}+2\pi\alpha^{\prime}F_{\mu\nu}= \nonumber\\
&\begin{pmatrix}
-\alpha^{\prime}\sqrt{\lambda}F(u_{b})^{-\frac{1}{2}}G(u_{b})^{\frac{\delta}{4}} & 0 & 0 & 2\pi\alpha^{\prime}E\\
0 & \alpha^{\prime}\sqrt{\lambda}F(u_{b})^{-\frac{1}{2}}G(u_{b})^{\frac{\delta}{4}} & 0 & 0\\
0 & 0 & \alpha^{\prime}\sqrt{\lambda}F(u_{b})^{-\frac{1}{2}}G(u_{b})^{\frac{\delta}{4}} & 0\\
-2\pi\alpha^{\prime}E & 0 & 0 & \alpha^{\prime}\sqrt{\lambda}F(u_{b})^{-\frac{1}{2}}G(u_{b})^{\frac{\delta}{4}}
\end{pmatrix}
\end{align}
Thus the DBI action becomes 
\begin{align}
\label{dbi}
\mathcal{S}_{\text{DBI}}=\frac{\alpha^{\prime}\lambda}{(2\pi)^{3}g_{s}}\int_{u=u_{b}} d^{4}x\; F(u_{b})^{-\frac{1}{2}}G(u_{b})^{\frac{\delta}{4}}\sqrt{F(u_{b})^{-1}G(u_{b})^{\frac{\delta}{4}}-
\left(\frac{2\pi E}{\sqrt{\lambda}}\right)^{2}}
\end{align}
\begin{figure}[tbp]
\centering 
\includegraphics[width=0.75\textwidth,height=0.5\textwidth,origin=c,angle=0]{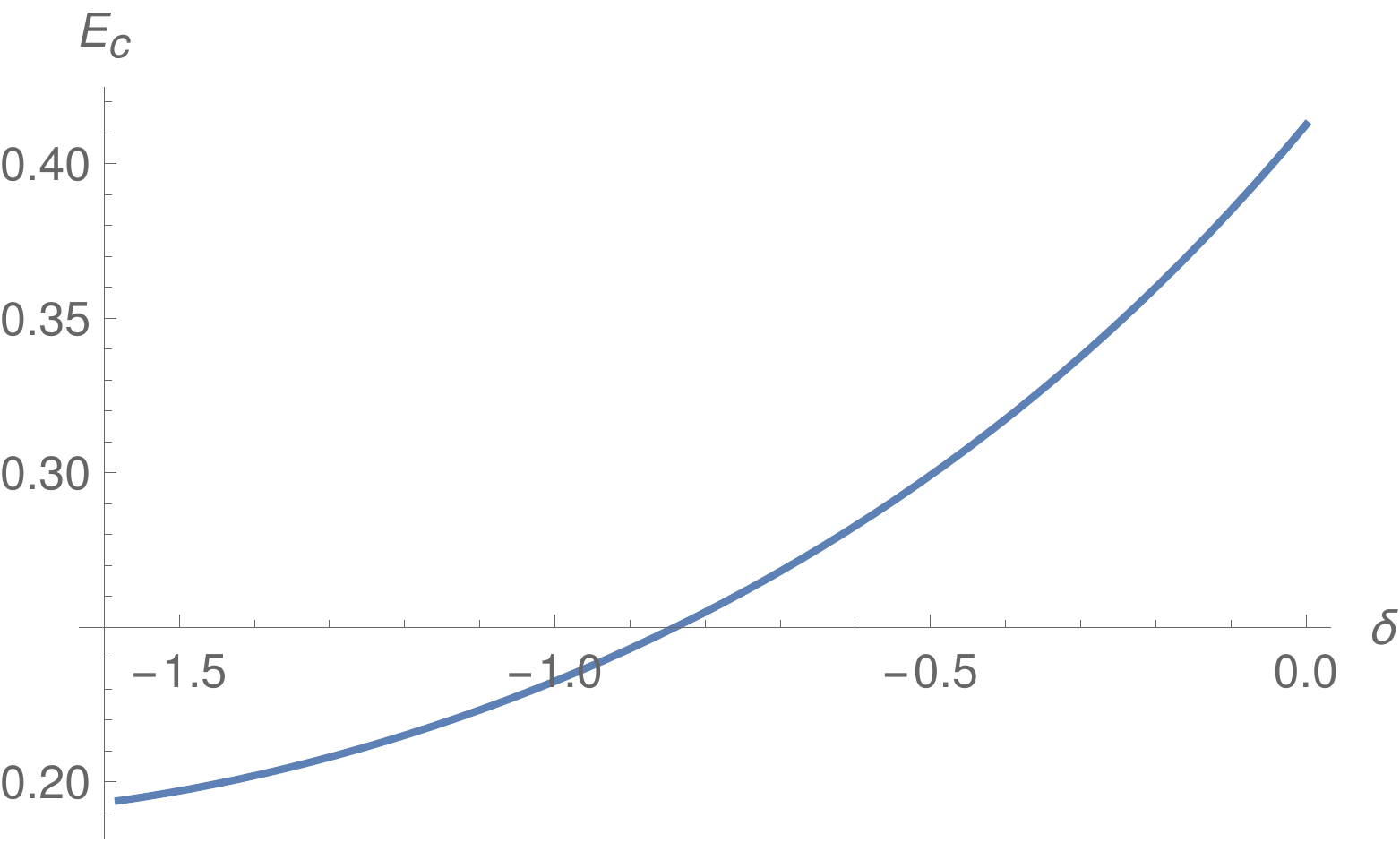}
\caption{This is the plot of critical electric field $E_{c}$ v/s the non-supersymmetric parameter $\delta$. In this figure we have set $\frac{\sqrt{\lambda}}{2\pi}=1$, $\frac{u_{0}}{u_{b}}=\frac{1}{0.5}$. 
According to this values the supersymmetric critical field would haven been 0.25.}
\label{fig2}
\end{figure}
Thus we see that \eqref{dbi} is not real for all values of the external electric field and there is an upper limit of the same. This limiting value is nothing but the critical electric field 
\begin{align}
 \label{npcritical}
 E_{c}=\frac{\sqrt{\lambda}}{2\pi}F(u_{b})^{-\frac{1}{2}}G(u_{b})^{\frac{\delta}{4}}
\end{align}
The functions $F(u)$ and $G(u)$ has been defined before \eqref{functions} . One can check that upto $\mathcal{O}\left(\frac{u_{0}}{u_{b}}\right)^{4}$, \eqref{npcritical} reduces
to \eqref{pcritical}. 
However in finding \eqref{npcritical} we have refrained from using perturbations of any sort and thus \eqref{npcritical} is the exact value. Let us check the behavior of it with
respect to the parameter $\delta$.

We see from Figure \ref{fig2} that the critical electric field is same as that of its supersymmetric cousin somewhere around $\delta=-0.85$, which matches more or less with our
perturbative analysis is the last section.
For values of $\delta>0.85$, the critical electric field is greater than the supersymmetric counterpart, and for $\delta<0.85$, the critical field is lesser. Thus the 
question raised in the last section is answered in the affirmative. The calculation in section \ref{section4} won't be affected drastically for small values of inter-quark
separation. (This is because it is the small separation behavior that decides the critical value.)

\section{Holographic Pair Production Rate for Non-supersymmetric Theories }
\label{section6}
In this section we calculate the pair production rate by using the method of circular Wilson loops. As indicated earlier in \eqref{exact} and \eqref{ideal}, to find the pair production rate, we need to find the 
on-shell value of the Nambu Goto action with string endpoints ending on a circular contour at the probe brane ($u=u_{b}$). For pure AdS the calculation of the same has been presented
in \cite{Semenoff:2011ng}\cite{Kawai:2015mha}\cite{Sato:2013pxa}. However, it is not possible to find exact solutions to the relevant equation of motions for the present case \eqref{throat}. Thus we will
resort to perturbative treatments like that of \cite{Allahbakhshi:2013rda} to calculate the circular Wilson loop and hence the decay rate to first order of the non-susy deformation parameter ($u_0^4$).
Since the metric \eqref{throat} enjoys circular symmetry we start by making an 
ansatz
\begin{align}
\label{guess}
x^0=r(\sigma)\cos \tau\;\;\;\;\;\;;\;\;\;\;\;x^3=r(\sigma)\sin \tau\;\;\;\;\;;\;\;\;\;\;\;u=u(\sigma)
\end{align}
In the above all other co-ordinates have been put to be constants as circular symmetry would imply. The parameter $\tau$ ranges from $(0,2\pi)$ while the parameter $\sigma$ is still arbitrary. Their
exists a diffeomorphism invariance of the Nambu-Goto action with which we can set $u=u(\sigma)$ to a function of our choosing. Putting the ansatz \eqref{guess} in \eqref{throat} we have the induced metric to be
\begin{align}
\label{induced}
 ds^2=\alpha^{\prime}\sqrt{\lambda}\Bigg[&\bigg(F(u)^{-\frac{1}{2}}G(u)^{\frac{\delta}{4}}\left(\frac{dr}{d\sigma}\right)^2+
 F(u)^{\frac{1}{2}}G(u)^{\frac{\delta-3}{4}}\left(\frac{du}{d\sigma}\right)^2\bigg)d\sigma^2+\nonumber\\
 &r^2F(u)^{-\frac{1}{2}}G(u)^{\frac{\delta}{4}}d\tau^2\Bigg]
\end{align}
From the above one can get the Nambu-Goto action to be of the form
\begin{align}
\label{ong}
 \mathcal{S}_{ng}=\frac{(\alpha^{\prime}\sqrt{\lambda})}{2\pi \alpha^{\prime}}\int_{0}^{2\pi}d\tau \int_{0}^{\sigma_b}d\sigma\;\sqrt{r^2 G(u)^\frac{\delta}{2}\bigg(F(u)^{-1}\left(r^{\prime}\right)^2+
 G(u)^{-\frac{3}{4}}\left(u^{\prime}\right)^2\bigg)}
\end{align}
For purposes of calculation we expand the function $F(u)$ and $G(u)$ in their leading order to the non-supersymmetric deformation parameter and we have,
\begin{align}
 \label{ongp}
 \mathcal{S}_{ng}&=\sqrt{\lambda} \int_{0}^{\sigma_b}d\sigma\;\sqrt{r^2\Big(1+\frac{\delta}{2}\frac{u_0^4}{u^4}\Big)\Bigg(\left(r^{\prime}\right)^2 u^4\Big(1+\frac{1}{2}\frac{u_0^4}{u^4}\Big)+
 \left(u^{\prime}\right)^2\Big(1-\frac{3}{4}\frac{u_0^4}{u^4}\Big)\Bigg)+\mathcal{O}\Big(\frac{u_o^8}{u^8}\Big)}\nonumber\\
 &\approx \sqrt{\lambda} \int_{0}^{\sigma_b}d\sigma\;\sqrt{r^2\Bigg(\left(r^{\prime}\right)^2 u^4\Big(1+A\frac{u_0^4}{u^4}\Big)+
 \left(u^{\prime}\right)^2\Big(1+B\frac{u_0^4}{u^4}\Big)\Bigg)}
\end{align}
Wherein
\begin{align}
 B=\frac{\delta+1}{2}\;\;\;\;\;;\;\;\;\;\;A=\frac{2\delta-3}{4}\;\;\;\;\;;\;\;\;\;\;A+\frac{5}{4}=B
 \end{align}
The above binomial expansion and all the others that follow is simply treating the Non-SuSy theory as a perturbation over the regular $\mathcal{N}=4$ SYM. In this paper we 
limit ourselves to first order perturbations.
Recall that we still had one diffeomorphism invariance left as mentioned before, with the help of which we set $\frac{du(\sigma)}{d\sigma}=1$. Thus \eqref{ongp} is simplified to
\begin{align}
 \label{ongpeff}
 \mathcal{S}_{ng}=\sqrt{\lambda}\int_{u_{t}}^{u_{b}} du\;\sqrt{r^2\bigg(\frac{dr}{du}\bigg)^2\Big(u^4+A u_0^4\Big)+\frac{r^2}{u^4}\Big(u^4+B u_0^4\Big)}
\end{align}
We would like to find out the function $r=r(u)$ which extremizes \eqref{ongpeff}. Extremizing the same one has to encounter the equation
\begin{align}
 \label{eqn}
  u^4&\big(u^4+Au_0^4\big)\bigg(\frac{d\rho}{du}\bigg)^2\bigg(2\big(u^4+Bu_0^4\big)-u^7\frac{d\rho}{du}\bigg)-4\rho\bigg(u^3\frac{d\rho}{du}\big(ABu_0^8+3Bu^4u_0^4+2u^8\big)\nonumber\\
  &-\big(u^4+Bu_0^4\big)^2\bigg)-2u^4\big(u^4+Au_0^4\big)\big(u^4+Bu_0^4\big)\rho\frac{d^2\rho}{du^2}=0
\end{align}
Where $\rho=r^2$. The above equation is very hard to solve in closed form. Thus we adopt perturbative techniques like that of \cite{Chowdhury:2019mqi}. To do so we decompose the 
solution to \eqref{eqn} as $\rho=\rho_{0}+u_0^4 \rho_1$ in which $\rho_0=-\frac{1}{u^2}$, and $\rho_1$ indicates the perturbation. From \eqref{eqn} the equation for $\rho_1$ 
to the leading order of $u_0^4$ is
\begin{align}
 \label{peqn}
2u^2\bigg(6(B-A)+2u^7\frac{d\rho_1}{du}+u^8\frac{d^2\rho_1}{du^2}\bigg)=0
 \end{align}
One can check that the above is solved by
\begin{align}
 \label{apparent}
 \rho_1(u)=\frac{A-B}{5u^6}+K^{\prime}
\end{align}
Thus the full solution is
\begin{align}
\label{profile}
 r^2(u)=\rho(u)= &u_0^4K^{\prime}-\frac{1}{u^2}+\frac{A-B}{5u^6}\frac{u_0^4}{u^6}\nonumber\\
 \equiv&  K-\frac{1}{u^2}+\frac{A-B}{5}\frac{u_0^4}{u^6}\nonumber\\
 =& K-\frac{1}{u^2}-\frac{1}{4}\frac{u_0^4}{u^6}
\end{align}
Where a redefinition of constant has been made. Now its time to relate the constant K to physical parameters. At $u=u_b$ the value of $r$ is the radius of the Wilson loop $R$.
Thus we have,
\begin{align}
\label{radius}
K=R^2+\frac{1}{u_b^2}-\frac{A-B}{5}\frac{u_0^4}{u_b^6}
\end{align}
From the above we can also find the value of the turning point $u_t$, since at the turning point the radius $r(u_t)=0$. Thus the equation which determines the turning point is, 
\begin{align}
\label{turning}
K=\frac{1}{u_t^2}\Big(1-\frac{A-B}{5}\frac{u_0^4}{u_t^4}\Big)
\end{align}
Now we proceed to calculate the on shell value of the Nambu Goto action \eqref{ongpeff} on the solution \eqref{profile}. We have,
\begin{align}
\label{osc}
 \mathcal{S}_{ng}=&\sqrt{\lambda}\int_{u_t}^{u_b}du\sqrt{\frac{1}{4}\bigg(\frac{d(r^2)}{du}\bigg)^2\big(u^4+Au_0^4\big)+(r^2)\Big(1+B\frac{u_0^4}{u^4}\Big)}\nonumber\\
 =&\sqrt{\lambda}\int_{u_t}^{u_b}du\sqrt{\frac{1}{4}\cdotp\frac{4}{u^6}\Big(1-\frac{3}{5}(A-B)\frac{u_0^4}{u^4}\Big)^2\big(u^4+Au_0^4\big)+\Big(1+B\frac{u_0^4}{u^4}\Big)\Big(K-\frac{1}{u^2}+\frac{A-B}{5}
 \frac{u_0^4}{u^6}\Big)}\nonumber\\
 =&\sqrt{\lambda}\int_{u_t}^{u_b}du\sqrt{K+K B\frac{u_0^4}{u^4}+\mathcal{O}
 \big(u_0^8\big)}
\end{align}
We neglect the $\mathcal{O}\big(u_0^8\big)$ term in the above. The integral of the remaining part cannot be done in closed form by using Mathematica. So we resort to perturbative methods 
again, and write
\begin{align}
\label{oscp}
\mathcal{S}_{ng}=&\sqrt{\lambda}\int_{u_t}^{u_b}\;du\sqrt{K}\bigg[1+\frac{B}{2}\frac{u_0^4}{u^4}+\mathcal{O}\big(u_0^8\big)\bigg]\nonumber\\
\approx&\sqrt{\lambda}\Bigg(\bigg[\sqrt{K}u\bigg]^{u_b}_{u_t}-\bigg[\frac{2B\sqrt{K}u_0^4}{u^3}\bigg]^{u_b}_{u_t}\Bigg)
\end{align}
So far so good, however the reader may agree that working with \eqref{oscp} is still daunting given that we now have to substitute the highly non-linear 
relations \eqref{radius},\eqref{turning} into it. Happily there is
a way out of this mess. Recall that our theme has been to work in the leading order of $u_0^4$ and the last two terms of \eqref{oscp} come with a $u_0^4$ of their own. Thus
to leading order we may substitute the usual AdS
relations (relating K to $u_t$ and $u_b$) in the last term of\eqref{oscp}, but use the Non-Susy relations \eqref{radius},\eqref{turning} in the first term of the same. Doing so we have
\begin{align}
\label{oscpp}
\mathcal{S}_{ng}=\sqrt{\lambda}&\Bigg(\;\sqrt{\big(R^2u_b^2+1\big)-\frac{A-B}{5}\frac{u_0^4}{u_b^4}}-\sqrt{1-\frac{A-B}{5}\frac{u_0^4}{u_t^4}}\nonumber\\
&+u_0^4\bigg[\frac{2B}{u_t^4}-\frac{2B\sqrt{R^2u_b^2+1}}{u_b^4}\bigg]\;\Bigg)\nonumber\\
\approx\sqrt{\lambda}&\Bigg(\;\sqrt{\big(R^2u_b^2+1\big)-\frac{A-B}{5}\frac{u_0^4}{u_b^4}}-\sqrt{1-\frac{A-B}{5}\frac{u_0^4}{u_b^4}\big(R^2u_b^2+1\big)^2}\nonumber\\
&+\frac{u_0^4}{u_b^4}\bigg[2B\big(R^2u_b^2+1\big)^2-2B\sqrt{R^2u_b^2+1}\bigg]\;\Bigg)\nonumber\\
\approx\sqrt{\lambda}&\Bigg(\;\sqrt{\big(R^2u_b^2+1\big)}-1+\frac{u_0^4}{u_b^4}\bigg[\frac{A+19B}{10}\big(R^2u_b^2+1\big)^2\nonumber\\
&-\frac{A-B}{10\sqrt{R^2u_b^2+1}}-2B\sqrt{R^2u_b^2+1}\bigg]\;\Bigg)
\end{align}
In the second line of the above we have used the usual AdS relations for the term $\frac{1}{u_t^4}$ to leading order as it is accompanied by a $u_0^4$. Again in the third line 
we have used a binomial expansion and retained
terms of leading order in $\frac{u_0^4}{u_b^4}$. 
Now, in presence of an electric field the effective action of the string has an extra piece, $S_B=T_0\int\;d\sigma d\tau B_{\mu\nu}\partial_{\sigma}x^{\mu}\partial_{\tau}x^{\nu}$. 
Specializing to constant electric field, the 
contribution of $S_B$ is a pure boundary term with  on-shell value $\pi R^2 E$, where R is the radius of the Wilson loop, $E=B_{01}$ and all other components of the $B_{\mu\nu}$ is set
to zero. The effective action is given by
\begin{align}
 \label{osceff}
 S_{eff}=&S_{ng}+S_{B}\nonumber\\
 =&\sqrt{\lambda}\Bigg(\sqrt{x}-1+\frac{u_0^4}{u_b^4}\bigg[\frac{A+19B}{10}x^2-\frac{A-B}{10\sqrt{x}}-2B\sqrt{x}\bigg]-\mathcal{E}x+\mathcal{E}\Bigg)
\end{align}
In the above, $x=R^2u_b^2+1$ and $E=\frac{\sqrt{\lambda}u_b^2}{\pi}\mathcal{E}$. Thus the radius R (x) is a free parameter in the expression \eqref{osceff}. Following \cite{Semenoff:2011ng},\cite{Bolognesi:2012gr}
the radius should be set to an an extremum of \eqref{osceff}. Instead of extremizing w.r.t. $R$, we extremize the action \eqref{osceff} w.r.t. the parameter $y=\sqrt{x}$. Doing so we find,
\begin{align}
 \label{extremize}
 0=\frac{dS_{eff}}{dy}=\sqrt{\lambda}\Bigg(1-2\mathcal{E}y+\frac{u_0^4}{u_b^4}\bigg[\frac{2(A+19B)}{5}y^3+\frac{A-B}{10y^2}-2B \bigg]\Bigg)
\end{align}
The radius $R$ should be set to be the solution of \eqref{extremize}, recall $y=\sqrt{R^2u_b^2+1}$. Thus the value of $y$ in the above equation is constrained and should always be greater than $1$.
This is because the 
radius of the Wilson loop should be a real number. A subtle point is that the range of parameter y should be restricted to half of the real line, because the radius R is nonnegative. The 
critical electric field $\mathcal{E}_c$ is the one for which the radius $R=0$, i.e. $y=1$. Setting so in the above we see,
\begin{align}
1-2\mathcal{E}_c+\frac{u_0^4}{u_b^4}\bigg[\frac{2(A+19B)}{5}+\frac{A-B}{10}-2B\bigg]=0
\end{align}
Thus
\begin{align}
 \label{ccritical}
E_c=&\frac{\sqrt{\lambda}}{2\pi}u_b^2\bigg[1+\frac{1}{8}\frac{u_0^4}{u_b^4}\Big(23+24 \delta \Big)\bigg]\nonumber\\
=& \frac{2\pi}{\sqrt{\lambda}}m_0^2\bigg[1+\frac{\lambda^2}{128\pi^4}\Big(23+24 \delta \Big)\frac{u_0^4}{m_0^4}\bigg]
\end{align}
We see that like \eqref{pcritical} and \eqref{npcritical} the value of the critical electric field is greater than he supersymmetric value for the value $\delta=-\frac{23}{24}$ and less than the
supersymmetric cousin otherwise.
Our perturbative analysis has even shown that the value of parameter $\delta$ for which this phase transition occurs is slightly bigger than -1 as can be seen from the non-perturbative DBI analysis. 
Now to find the expression of the pair production rate we need to solve \eqref{extremize} for y. As can be seen, that is not analytically possible. We thus resort to perturbative treatments again and write,
\begin{align}
\label{a}
y=y_0+\frac{u_0^4}{u_b^4}y_1
\end{align}
$y_0$ being the usual AdS solution i.e. $u_0=0$ in \eqref{extremize}. The value of $y_0$ is $\frac{1}{2\mathcal{E}}$. We put the above relation in the equation in \eqref{extremize} to get upto leading order in
$\frac{u_0^4}{u_b^4}$.
\begin{align}
\label{b}
 y_{1}=\frac{1}{2\mathcal{E}}\bigg[\frac{A+19B}{20}\frac{1}{\mathcal{E}^3}+\frac{2(A-B)}{5}\mathcal{E}^2-2B\bigg]
\end{align}
Now we put \eqref{a} and \eqref{b} in \eqref{osceff}, i.e. find out the on-shell action. Retaining terms in leading order of $\frac{u_0^4}{u_b^4}$ leads us to.
\begin{align}
 \label{actiononshell}
 S_{eff}^{onshell}=&\frac{\sqrt{\lambda}}{2}\Bigg(\frac{1}{2\mathcal{E}}-2+2\mathcal{E}+\frac{u_0^4}{u_b^4}\bigg[\frac{A+19B}{80}\frac{1}{\mathcal{E}^4}+\frac{B-A}{10}\mathcal{E}-\frac{2B}{\mathcal{E}}\bigg]\Bigg)
 \nonumber\\
 &=\frac{\sqrt{\lambda}}{2}\Bigg(\frac{1}{2\mathcal{E}}-2+2\mathcal{E}+\frac{u_0^4}{u_b^4}\bigg[\frac{40\delta+39}{320}\frac{1}{\mathcal{E}^4}+\frac{1}{40}\mathcal{E}-\frac{\delta+1}{\mathcal{E}}\bigg]\Bigg)
\end{align}
The pair production rate of quark anti-quark pairs per unit volume per unit time is given by the formula, $\Gamma \sim e^{-S_{eff}^{onshell}}$. Note that we are using reduced parameter $\mathcal{E}=
\frac{\pi}{\sqrt{\lambda}u_{b}^2}E$, in terms of
which the pure AdS pair production rate(per unit volume per unit time) is given by \cite{Semenoff:2011ng}\cite{Bolognesi:2012gr}\cite{Sato:2013pxa}.
\begin{align}
\label{pureproductionrate}
 \Gamma_{susy}\sim\text{exp}\Bigg[-\frac{\sqrt{\lambda}}{2}\bigg(\sqrt{\frac{\frac{1}{2}}{\mathcal{E}}}-\sqrt{\frac{\mathcal{E}}{\frac{1}{2}}}\bigg)^2\Bigg]
\end{align}
For the pure AdS/Supersymmetric scenario, the critical electric field is $\mathcal{E}_c=\frac{1}{2}$ i.e. $E_c=\sqrt{\lambda}\frac{u_b^2}{2\pi}$. Now unlike the supersymmetric case, 
the pair production rate cannot be 
brought in closed form. We will have to resort to numerical calculations. We present the plots of pair production rate. Computation of the fluctuation prefactor (i.e. the $\frac{(eE)^2}{(2\pi)^3}$ term 
in \eqref{fluc}) 
is somewhat a open question in holography, which is the reason we have plotted $e^{-S_{onshell}}$ instead of $\Gamma$.
\begin{figure}[tbp]
\centering 
\includegraphics[width=0.45\textwidth,height=0.40\textwidth,origin=c,angle=0]{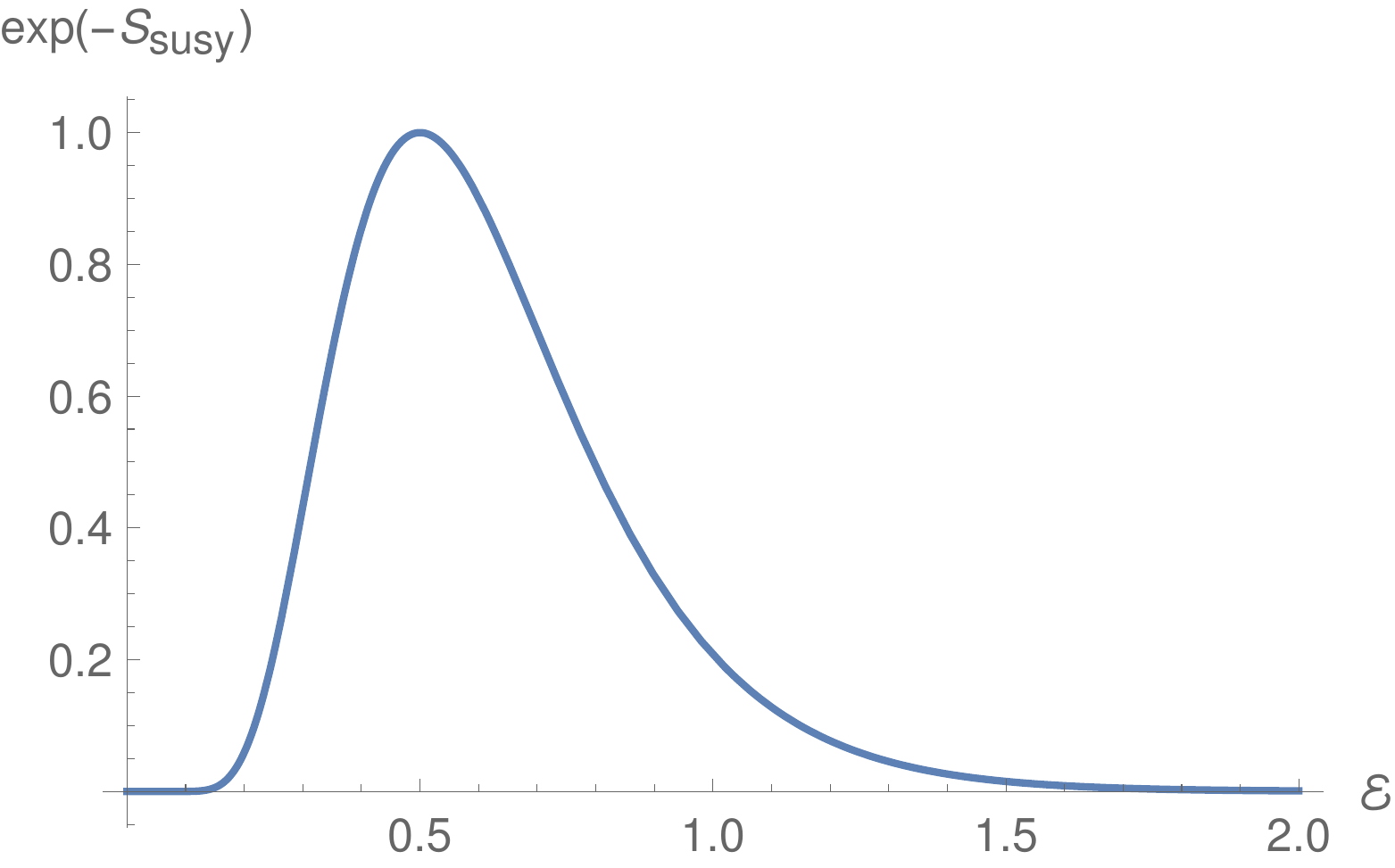}
\includegraphics[width=0.45\textwidth,height=0.40\textwidth,origin=c,angle=0]{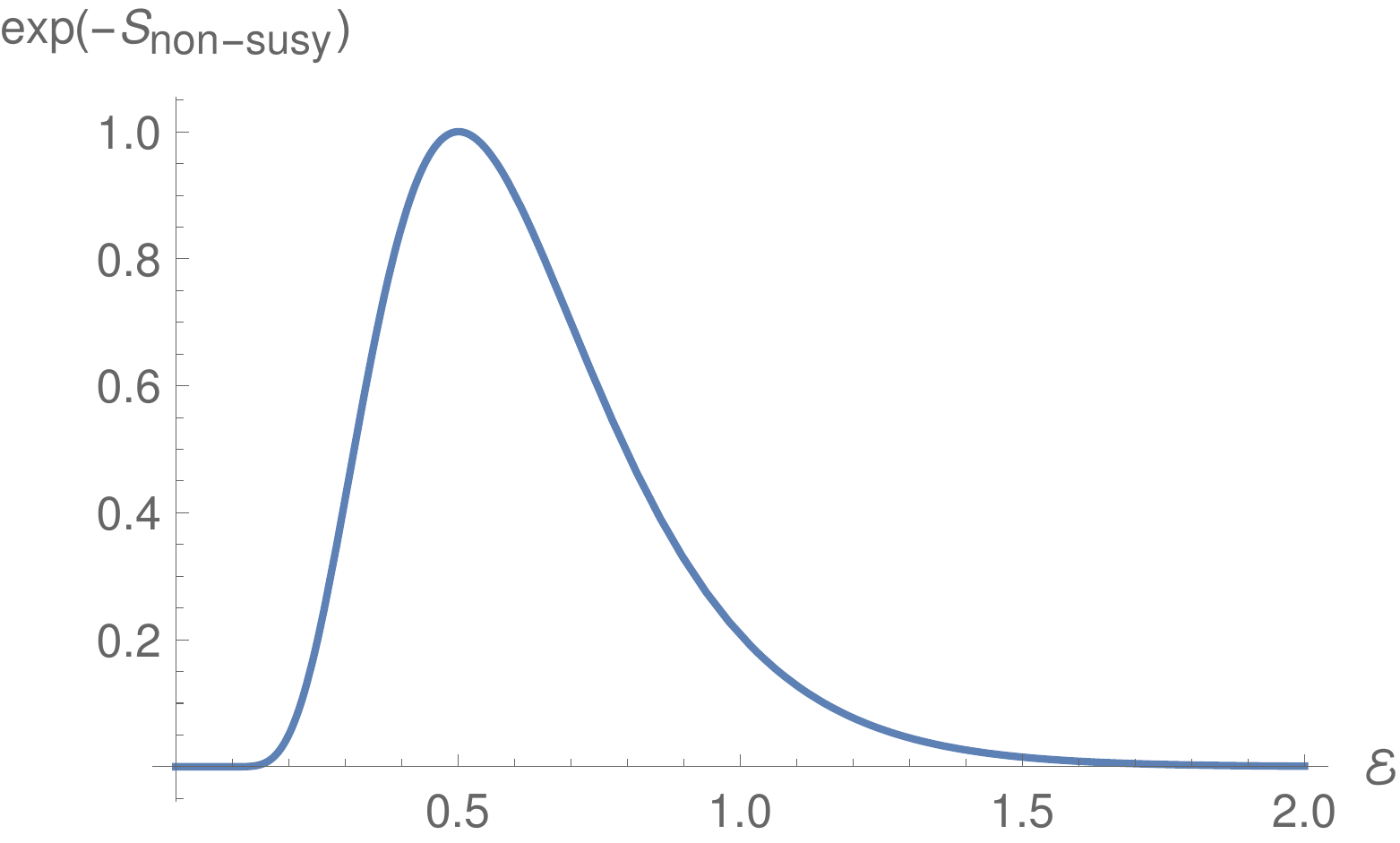}
\caption{The figure on the left illustrates relation between the pair production rate and applied electric field for pure $\mathcal{N}=4$ SYM \eqref{pureproductionrate}. In our units the critical 
electric field is at $\mathcal{E}_c=\frac{1}{2}$. The right figure is for non-supersymmetric case \eqref{actiononshell} for the value $\delta=-0.75$. Almost no change in the profile is found ! }
\label{fig34}
\end{figure}
\begin{figure}[tbp]
\centering 
\includegraphics[width=0.45\textwidth,height=0.40\textwidth,origin=c,angle=0]{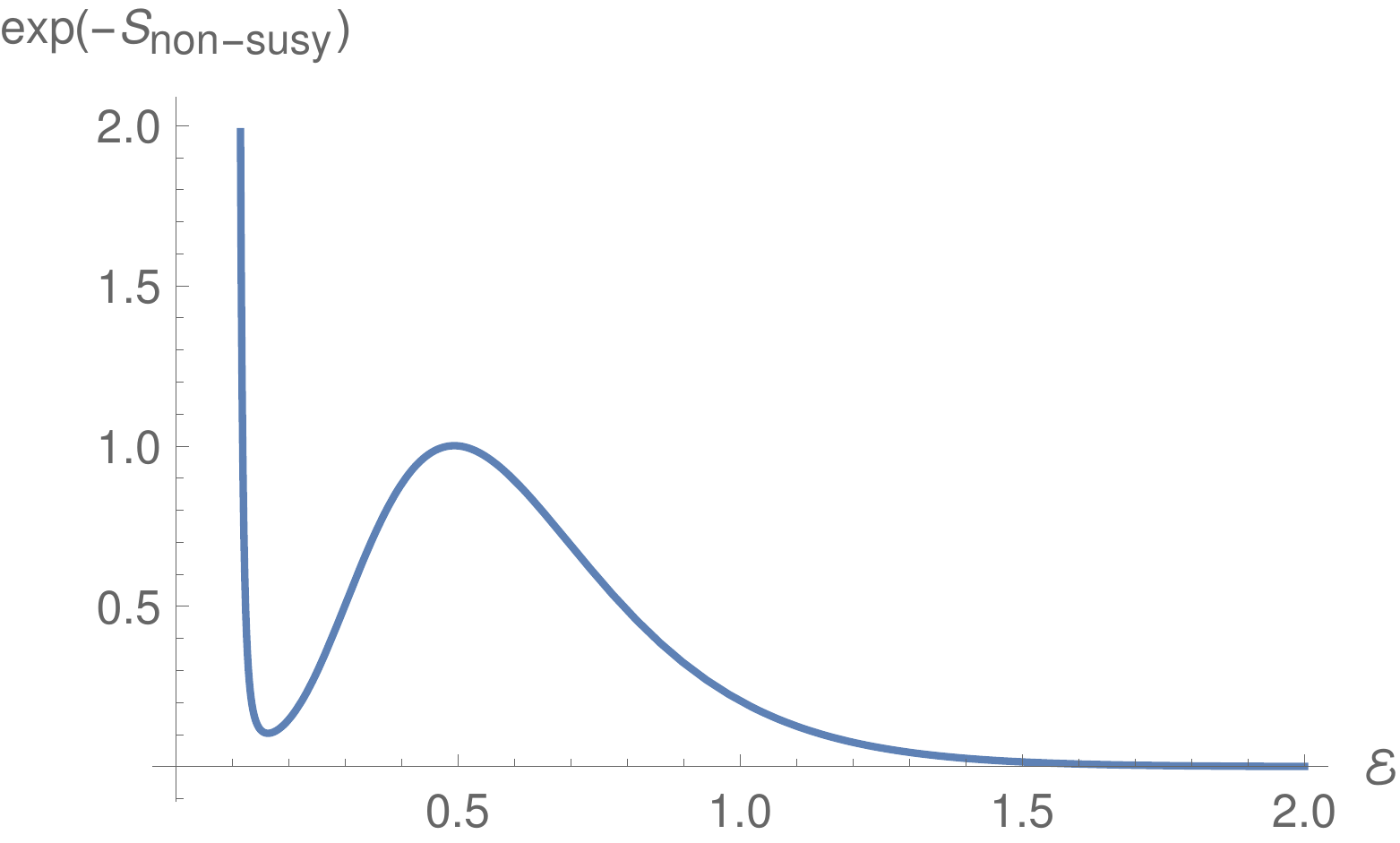}
\includegraphics[width=0.45\textwidth,height=0.40\textwidth,origin=c,angle=0]{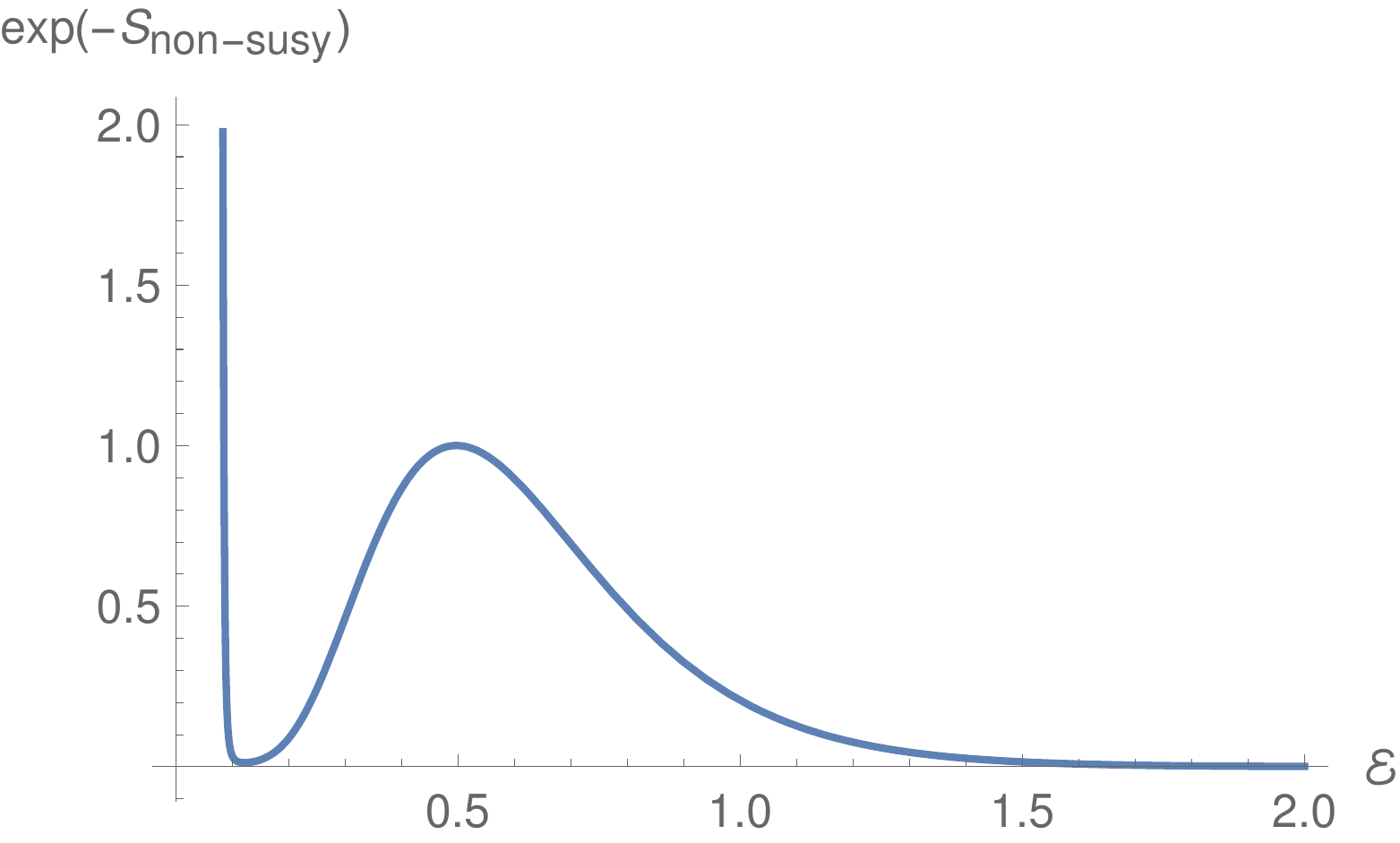}
\caption{We plot the pair production rate for non-supersymmetric Yang Mills for the value $\delta=-\sqrt{2}$. On the right we have $\big(\frac{u_0}{u_b}\big)^4=\big(\frac{1}{5}\big)^4$ and on the left
,$\big(\frac{u_0}{u_b}\big)^4=\big(\frac{1}{3.75}\big)^4$.}
\label{fig56}
\end{figure}

Physical interpretation of the plots : As commented in the caption, no stark contrast is found between the susy and non-susy case in Figure \ref{fig34}. This is the case when the parameter is greater than 
(somewhere around) $-0.975$. The reason for this can be seen from \eqref{actiononshell}. For low electric field the pair production rate is dominated by the 
$\frac{u_0^4}{u_b^4}\frac{40\delta+39}{320}\frac{1}{\mathcal{E}^4}$ term. Above $\delta=-0.975$ the effective correction at the low electric field limit is positive. At the limit of high electric field limit the
correction of pair production rate due to the non-susy deformation parameter is always positive. This is reason that the high electric field limit, the behavior of non-susy pair production
rate is same as its supersymmetric cousin for all values of parameters. Startling effects happen when the parameter $\delta<-0.975$, for which the 
plot is shown in Figure \ref{fig56}. Let us recall, the prefactor of the pair production rate is given by field theoretic calculations to be $\frac{(eE)^2}{(2\pi)^3}$, see \eqref{fluc}. Although the 
holographic calculation of the fluctuation prefactor is currently a mystery, it
should definitely match with field theoretic calculation for low electric field. For small applied electric fields the production rate shoots up signaling in \textit{non-perturbative instability of the vacuum}. 
We say "non-perturbative" because Schwinger effect is by itself a
non-perturbative phenomenon.  We see that the limit $\delta=-0.975$ apprx. is for more interesting than 
earlier imagined. 
\\
\\
Let us end by writing down the pair production rate per unit spatial volume per unit time for non-susy Yang Mills.
\begin{align}
 \label{finalrate}
 \Gamma_{\text{non-susy}}\approx \text{exp}\Bigg[-\frac{\sqrt{\lambda}}{2}\Bigg(&\frac{2\pi m_0^2}{\sqrt{\lambda}}\frac{1}{E}-2+\frac{\sqrt{\lambda}}{2\pi m_0^2}E+\frac{\lambda^2u_0^4}{16\pi^4 m_0^4}
 \bigg\{\frac{4\pi^4m_0^8\big(40\delta+39\big)}{5\lambda^2}\frac{1}{E^4}\nonumber\\
 &+\frac{\sqrt{\lambda}}{160\pi m_0^2}E-\frac{4\pi m_0^2\big(\delta+1\big)}{\sqrt{\lambda}}\frac{1}{E}\bigg\}\Bigg)\Bigg]
\end{align}

\section{Conclusion}
\label{section7}
In this paper we have studied pair production (Schwinger Effect) in presence of external electric field for Non-susy AdS/CFT using three methods in the literature. In section \ref{section4} we have 
done a potential analysis by calculating rectangular Wilson loops and have 
analytically calculated the critical electric field below which pair production happens via a tunneling phenomenon (and above which the quark-antiquark potential ceases to put up a potential barrier.
We have seen that the
critical electric field is higher / lower than its supersymmetric counterpart depending on the value of the non-supersymmetric parameter $\delta$(in that section we have used a metric perturbation to 
ease up the calculation).
We have also confirmed the same from the DBI analysis of the critical 
electric field in section \ref{section5} where no such approximation has been made. As can be seen from the main body of work, the correction absence of supersymmetry
yields on the critical electric field (w.r.t. its supersymmetric value) is neither positive nor negative definite\eqref{pcritical}\eqref{npcritical}\eqref{ccritical}. Thus such a relation cannot be
conveniently used to be a indirect evidence for presence or absence of supersymmetry since the modulus of the correction is parameter dependent. Also note that from Figure \ref{fig9} no trace of confinement is seen
contrary to earlier findings in literature
 (for confinement some of the plot should have been positive with non negative slope all along). Next in section \ref{section6} we have performed the 
analysis for pair production rate for quark-antiquark pairs using 
circular Wilson loops. Since the relevant 
equations are rather impossible to solve, we have resorted to
perturbative analysis, which can be thought of as perturbation over $\mathcal{N}=4$ SYM by a supersymmetry breaking term with coupling constant proportional to $\big(\frac{u_0}{u_b}\big)^4$ parametrized
by $\delta$. We
have explicitly found out the profile for circular Wilson loop for Non-susy AdS/CFT upto first order of $u_0^4 $. To our knowledge this is the first time such a solution has been obtained. We proceed to
find the on-shell
value of the Nambu-Goto action on the profile found and relate it to pair production rate. We see for a regime of allowed value of the parameter $\delta$ the pair production rate shoots up as external 
electric field  \textit{decreases 
towards zero}. In stark contrast to confinement (as earlier reported) this signals that the \textit{vacuum of the dual Non-SuSy gauge theory is non-perturbatively unstable for the regime of 
the parameter $-\sqrt{\frac{5}{2}}\leqslant\delta<-\frac{39}{40}$} (apprx) (a confining potential would show the exact opposite). Thus the \textbf{field theory dual of the Non-SUSY geometry considered here is not
confining but is the exact opposite i.e non-perturbatively unstable}.
A relevant question is to
find this instability from potential analysis and DBI analysis, something which eludes us at this moment.


\bibliographystyle{JHEP}
\bibliography{NSuhse}
\end{document}